%% file: main.tex
\documentclass[12pt]{iopart}
\usepackage{xcolor}
\usepackage{graphicx}
\usepackage{dcolumn}
\usepackage{bm}
\usepackage{multirow}
\usepackage[numbers,sort&compress,square]{natbib}
\usepackage[numbers,sort&compress,square]{natbib}
\usepackage{hyperref}
\usepackage{subcaption}
\usepackage{verbatim}
\usepackage{soul}
\usepackage{iopams} 
\usepackage{acronym}
\usepackage[mathlines,running]{lineno}
\usepackage{tabularx}
\usepackage{listings}

\newcommand{\LSU}{Department of Physics, Louisiana State University, 202 Nicholson Hall
Baton Rouge, LA 70803, USA}
\newcommand{\MIT}{LIGO Laboratory, Massachusetts Institute of Technology, Cambridge, MA 02139, USA}
\newcommand{\LLO}{LIGO Livingston Observatory, Livingston, LA 70754, USA}
\newcommand{\LHO}{LIGO Hanford Observatory, Richland, WA 99352, USA}
\newcommand{\CIT}{LIGO Laboratory, California Institute of Technology, Pasadena, CA 91125, USA}

\usepackage[mathlines,running]{lineno}
\begin{document}

\title{Modeling and Reduction of High Frequency Scatter Noise at LIGO Livingston.}
\author{Siddharth Soni$^{1}$, 
Jane Glanzer$^{2}$, 
Anamaria Effler$^{3}$, Valera Frolov$^{3}$, Gabriela Gonz\'alez$^{2}$, Arnaud Pele$^{4}$, Robert Schofield$^{5}$}

\address{$^1$\MIT}
\address{$^2$\LSU}
\address{$^3$\LLO}
\address{$^4$\CIT}
\address{$^5$\LHO}
\begin{abstract}
The sensitivity of aLIGO detectors is adversely affected by the presence of noise caused by light scattering. Low frequency seismic disturbances can create higher frequency scattering noise adversely impacting the frequency band in which we detect gravitational waves. In this paper, we analyze instances of a type of scattered light noise we call ``Fast Scatter" that is produced by motion at frequencies greater than 1 Hz, to locate surfaces in the detector that may be responsible for the noise. 
We model the phase noise to better understand the relationship between increases in seismic noise near the site and the resulting Fast Scatter observed. We find that mechanical damping of the Arm Cavity Baffles (ACBs) led to a significant reduction of this noise in recent data. For a similar degree of seismic motion in the  $1$--$3~\mathrm{Hz}$ range, the rate of noise transients is reduced by a factor of $\sim$ 50.

\end{abstract}

\section{Introduction}\label{section_intro}

Transient noise is a common occurrence in the Advanced LIGO (aLIGO) and Advanced Virgo (AdV) gravitational wave (GW) detectors \cite{TheLIGOScientific:2014jea, TheVirgo:2014hva}.
Between 2015 and 2020, three Observing runs were completed that led to the detection of 90 gravitational wave events \cite{abbott2019gwtc, LIGOScientific:2021usb, LIGOScientific:2021djp}.
The fourth Observing run (O4) began on May 24 2023 with LIGO and KAGRA detectors resuming the search for gravitational waves. Addition of new technologies and multiple upgrades including but not limited to frequency dependent squeezing, new test masses, increased laser power, and reduced low frequency noise contributed to increased sensitivity of the LIGO detectors in O4 \cite{cahillane2022review, PhysRevLett.124.171102}.

Most GW events are very short in duration, meaning that their signal needs to be extracted from very large populations of transient glitches \cite{abbott2020guide}. Noise transients in the data can mask the compact binary coalescence (CBC) signals, adversely impacting the parameter estimation \cite{powell2018parameter}. Another complication with transient noise is that it can lead to false alerts by the search pipelines. During the third Observing run (O3), 23 out of the 80 low latency alerts were retracted as their origin was deemed instrumental or environmental artifacts \cite{Soni:2023pxg, LIGOScientific:2021djp, LIGOScientific:2021usb}. 
Understanding this noise and its impact on the detector is crucial for determining the astrophysical nature of the candidate event \cite{abbott2016characterization}.

Multiple detector characterization tools are used to detect and classify transient events. Specifically, Omicron and Gravity Spy are used in this analysis \cite{robinet2020omicron, Robinet:2015om, Robinet:2020lbf, zevin2017gravity}. Omicron is an algorithm that makes use of a Q transform to search for excess power in LIGO data.  From this algorithm, triggers are generated and given a set of parameters, such as the event time, amplitude, frequency, duration etc. The classification of triggers is done with Gravity Spy. Gravity Spy is a machine learning project that uses convolutional neural networks to classify transient noise events based on their  glitch morphology ~\cite{Zevin:2016qwy,Bahaadini:2018git,Coughlin:2019ref,Soni:2021cjy, glanzer2022data}. Other tools including Hveto, Lasso, iDQ and the Detector Characterization Summary pages are used to study noise correlations between the primary gravitational wave channel and different detector components \cite{smith2011hierarchical, walker2018identifying, essick2020idq, davis2021ligo}.

Scattered light is one of several types of noise sources present in the LIGO detectors. It occurs when a small fraction of stray light strikes a moving surface, gets reflected back towards the point of scattering and rejoins the main laser beam. This stray light thus introduces a time dependent phase modulation $\delta \phi_{sc}(t)$ to the static phase $\phi_{0}$ of the main beam. This gives rise to phase noise $h_{ph}(f)$:

\begin{equation}
    {\phi}(t)  = {\phi}_{0} + {\delta\phi_{sc}(t)} =  \frac{4{\pi}}{{\lambda}}\left|{ x_{0} + {\delta} x_{sc}(t)}\right| \label{eq:1}
\end{equation}

\begin{equation}
    h_{ph}(f) = A\frac{{\lambda}}{8{\pi}L}\mathcal{F}\left[\sin{\delta{\phi}(t)}\right] \label{eq:2}
\end{equation}
This phase shifted field builds up in the arms due to arm cavity gain $\Gamma$ and results into radiation pressure or amplitude noise $h_{rad}$:
\begin{equation}
    h_{rad}(f) = A\frac{2{\Gamma }P}{MLc}\frac{2}{\Omega^2 - \omega^2} \mathcal{F}\left[\cos{\delta{\phi}(t)}\right] \label{rad_eqn}
\end{equation}
$\mathcal{F}$ represents the fourier transform, ${\Gamma}$ $= 13.58$ here is cavity signal gain, M $= 40$ kg is mirror mass, P is arm power, $c$ is speed of light and ${\Omega}$ is the suspension eigenfrequency~\cite{Ottaway:12}.
${\lambda}$ is the laser wavelength, $x_{0}$ is the static path that corresponds to the static phase $\phi_{0}$ while  $\delta x_{sc}$ is the time-dependent displacement of the scattering surface which gives rise to additional phase $\delta \phi_{sc},(t)$, A is the stray light amplitude that couples to the main beam and L is the arm length (4 kms) \cite{Accadia:2010zzb, Ottaway:12}.
If the scattering surface motion $\delta x_{sc}$ becomes a significant fraction of, or greater than $\lambda$,
\textit{fringe wrapping} occurs and the low frequency ground motion gives rise to high frequency noise in h(t). This phenomenon is known as \textit{upconversion} \cite{corey_thesis}. The phase noise $h_{ph}(f)$ appears as arches in the time-frequency spectrograms of the gravitational wave channel \cite{Chatterji:2004qg, gwdetchar}. Surface irregularities in the optics and the Gaussian tail of the main laser beam are two primary sources of stray light.

Differentiating both sides of Eq \ref{eq:1} leads to the peak or maximum frequency of the noise:
\begin{equation}
    f(t) = \frac{\left|2 n v(t)\right|}{\lambda} \label{eq:3}
\end{equation}
n represents nth harmonic in case of multiple reflection from the scattering surface, v is the velocity of the scatterer.

During O3, two separate populations of scattering transients were observed. Along with the prototypical long duration scattering arches present in earlier observing runs known as Slow Scattering, we noticed the presence of short duration scattering arches in O3. This population was named ``Fast Scattering"  due to its high frequency arches \cite{alog:josh_andy_fs, Soni:2021cjy}.
In the time-frequency spectrograms, slow scattering arches have a typical duration of 1 second or more, whereas fast scattering arches are much shorter in duration ($\sim$ 0.2 secs).
Reaction Chain (RC) tracking employed during O3b led to reduction in the rate of Slow Scatter \cite{Soni_2021}. 
Some version of high frequency scatter, known as ``Scratchy" was present during the second Observing run (O2) and was fixed by damping the motion of swiss cheese baffles \cite{alog:swiss_cheese_robert}.
There have been multiple efforts to characterize and reduce scattered light noise coupling in aLIGO and AdV \cite{longo2021daily, wkas2022high, valdes2017hilbert,Was:2020ziy, tolley2023archenemy, udall2022bayesian, Meinders:2015qxa}.

\begin{figure}
\captionsetup[subfigure]{font=scriptsize,labelfont=scriptsize}
   \centering
    \begin{subfigure}[b]{0.45\textwidth}
        \centering
         \includegraphics[width= \textwidth]{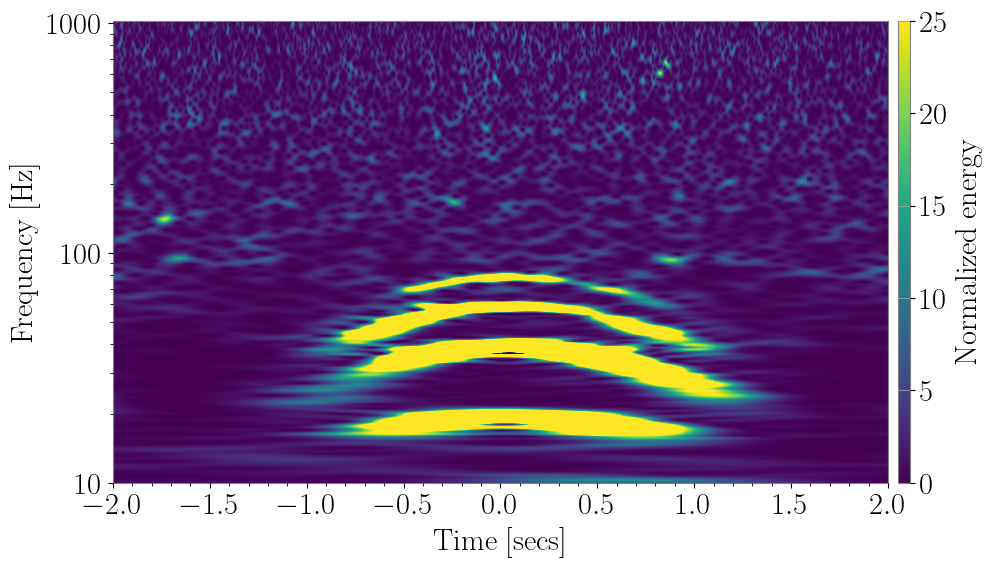}
         \caption{Slow Scattering arches.}
         \label{fig:slowscat}
    \end{subfigure}
    \hfill
    \begin{subfigure}[b]{0.45\textwidth}
        \centering
         \includegraphics[width =\textwidth]{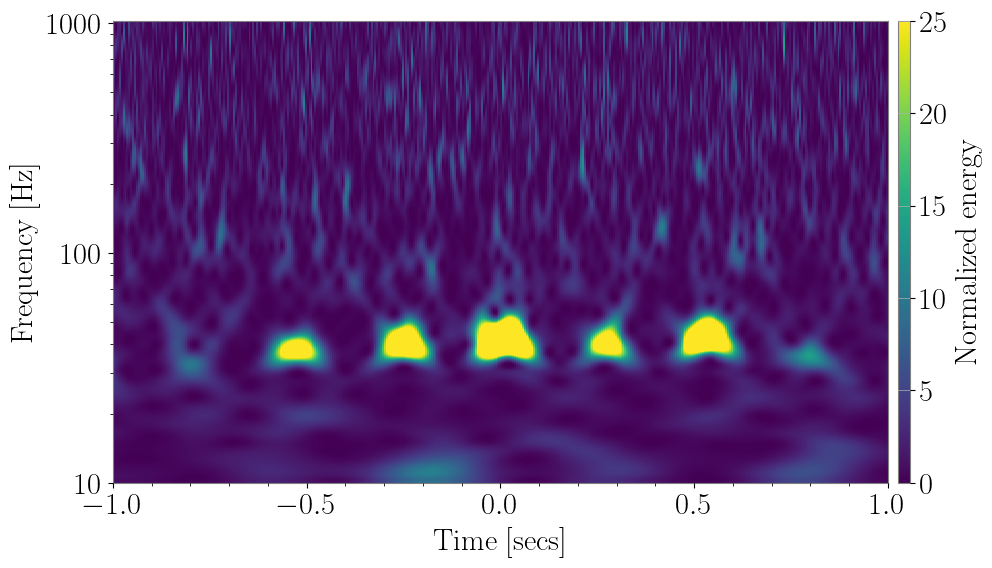}
          \caption{Fast Scattering arches.}
         \label{fig:fastscatter}

    \end{subfigure}
    \caption{Time-frequency spectrograms of transient noise in h(t) due to scattered light. \emph{Left} shows Slow scattering arches and \emph{right} shows Fast scattering.}
    \label{fig:slow_fast}
    
\end{figure}

Figure  \ref{fig:slow_fast} shows a spectrogram comparison between the two types of scatter observed in LIGO detector during O3. 
Scattering noise is intimately linked to movement of the ground near the detector. Ground motion is measured along the X, Y and Z axes using seismometers located at the End and Corner stations in the LIGO detector. The raw measurement is then bandpassed in several frequency bands between $0.03 ~\mathrm{Hz}$ and $30~\mathrm{Hz}$ and plotted on the LIGO Summary Pages \cite{duncan_macleod_2021_4975045}. The rate of Slow Scatter is correlated with ground motion in the earthquake band $0.03$--$0.1~\mathrm{Hz}$ and the microseismic band $0.1$--$0.3~\mathrm{Hz}$. The rate of Fast Scatter on the other hand, has been found to be correlated with the ground motion in the microseismic band $0.1$--$0.3~\mathrm{Hz}$ and the anthropogenic band $1$--$6~\mathrm{Hz}$ \cite{dcc_LVK_mar_2023_scatter, Soni_2021}. For more details on slow scattering and its reduction, we refer the readers to \cite{Soni_2021}. In this document, we focus on fast scattering at LIGO Livingston (LLO) during O3.

\section{Fast Scattering in O3 and Post O3 data}\label{sec:two}
As shown in Figure \ref{fig:slow_fast}, fast scattering occurs as short duration arches (as opposed to the long duration slow scattering arches) in the time--frequency spectrograms of the primary GW channel $h(t)$. These transients  usually impact the $h(t)$ sensitivity between $10~\mathrm{Hz}$ and $100~\mathrm{Hz}$ but occasionally can go as high as $400~\mathrm{Hz}$ \cite{alog:dec_15_2020_fs}.
Mainly due to differences in the anthropogenic ground motion near the site, as well as the higher sensitivity in $10-60~\mathrm{Hz}$ band between LLO and LIGO Hanford (LHO), fast scattering is a lot more noticeable at LLO \cite{alog:ashley_fs_rate, buikema2020sensitivity}. 

    
\begin{figure}[ht]
\captionsetup[subfigure]{font=scriptsize,labelfont=scriptsize}
   \centering
    \begin{subfigure}[b]{\textwidth}
        \centering
         \includegraphics[width= \textwidth]{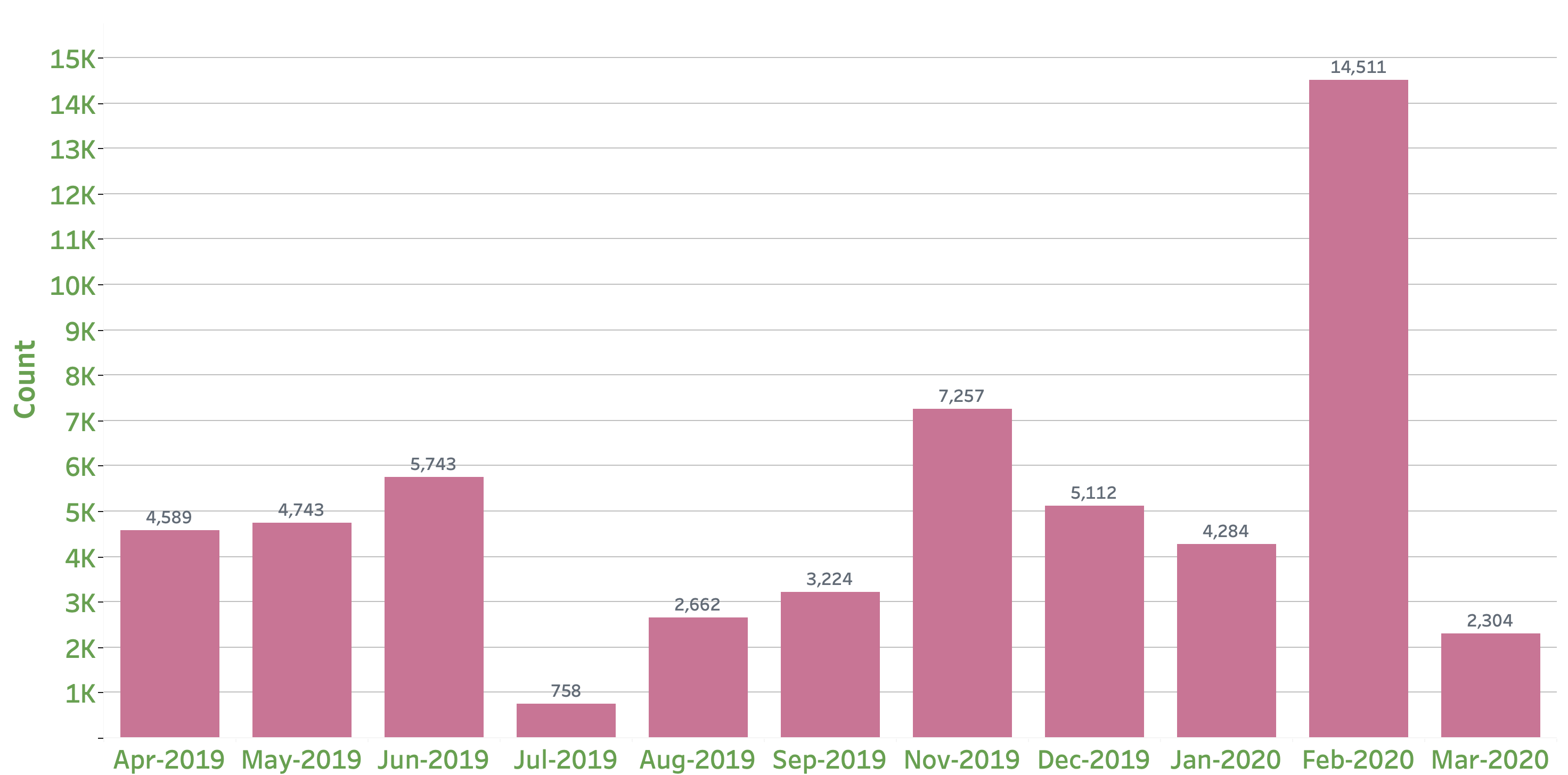}
         
    \end{subfigure}
    \caption{Fast Scattering glitches during each month of O3, as classified by Gravity Spy above a confidence of $90 \%$. High microseism and the reduction in Slow Scatter after implementation of RC tracking lead to increased visibility and rate of Fast Scatter in Feb 2020.}
    \label{fig:fsO3}
    
\end{figure}

Fast scatter was the most frequent source of transient noise at LLO during O3. The average rate of these transients in O3 was 9 per hour. At Hanford on the other hand, the rate of fast scatter transients was only $0.21$ per hour. At LLO, about $27 \%$ of all the O3 glitches were classified as ``Fast Scattering" by Gravity Spy with a confidence of 90\% or more \cite{glanzer2022data}. Figure \ref{fig:fsO3} shows the montly distribution of these transients in O3 at LLO. 
These glitches have been observed to occur when there is increased ground motion in both the microseismic $0.1$--$0.3~\mathrm{Hz}$ and anthropogenic $1$--$6~\mathrm{Hz}$ frequency bands. Ocean waves, winds, thunderstorms, human activity near the site, logging, construction activity and trains near the Y End station of LLO are the primary causes of ground motion in these frequency bands, impacting the h(t) sensitivity \cite{dcc:noise-sprint, microsiesm_paper}.

\begin{figure}[ht]
    \centering
    \includegraphics[width=\textwidth]{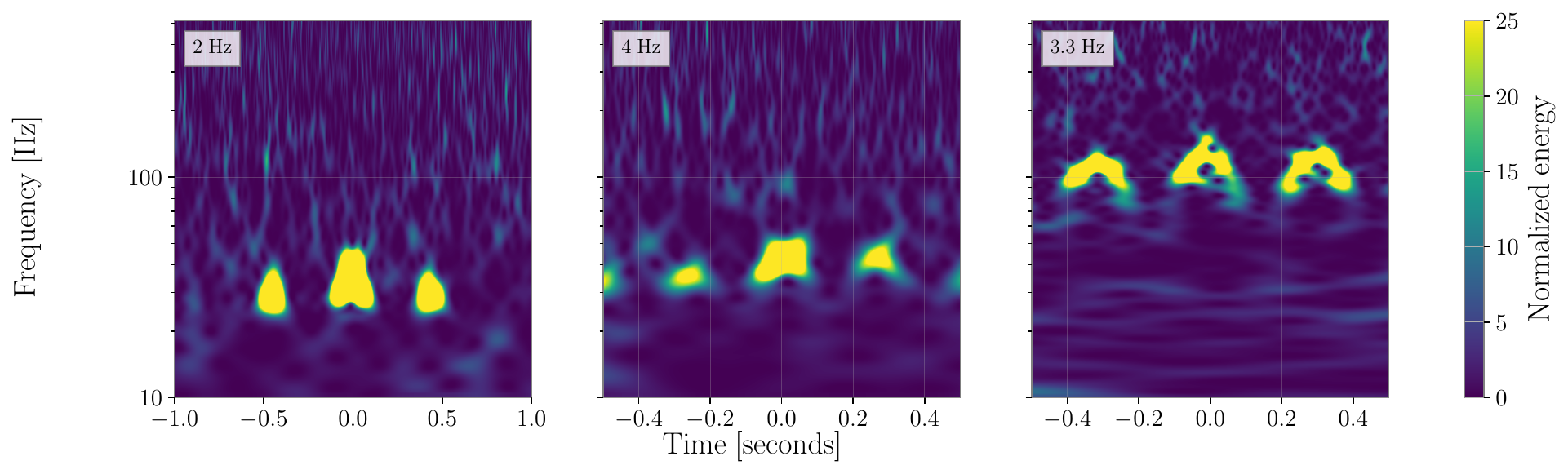}
    \caption{Time-frequency spectrograms of the different types of Fast Scatter that has been observed. In the 2 Hz fast scatter, the arches are separated by $0.5$ secs whereas in 4 Hz, the arch separation is $0.25$ secs. Both of these population were observed during O3, whereas 3.3 Hz fast scatter is present in the post O3 data.}
    \label{fig:fast_population}
\end{figure}

There have been several different types of fast scatter observed in the gravitational wave data channel, chief among them are 2 Hz, 3.3 Hz, and 4 Hz Fast Scatter. Figure~\ref{fig:fast_population} compares these different populations. Next we discuss our investigations into these different sub-populations and their detector couplings. 

\subsection{4 Hz Fast Scatter}
The 4 Hz sub-population of fast scatter consists of arches separated by $\sim$ $0.25~\mathrm{sec}$ as shown in Figure~\ref{fig:fast_population}. This is also the most common type of fast scatter during O3. The rate of 4 Hz noise has been found to be correlated with an increase in anthropogenic ground motion. Human activity, trains, bad weather conditions, road work near the site all contribute to an increased rate \cite{glanzer2023noise,Soni:2021cjy}. 

\begin{figure}[ht]
\captionsetup[subfigure]{font=scriptsize,labelfont=scriptsize}
   \centering
    \begin{subfigure}[b]{0.48\textwidth}
        \centering
         \includegraphics[width= \textwidth]{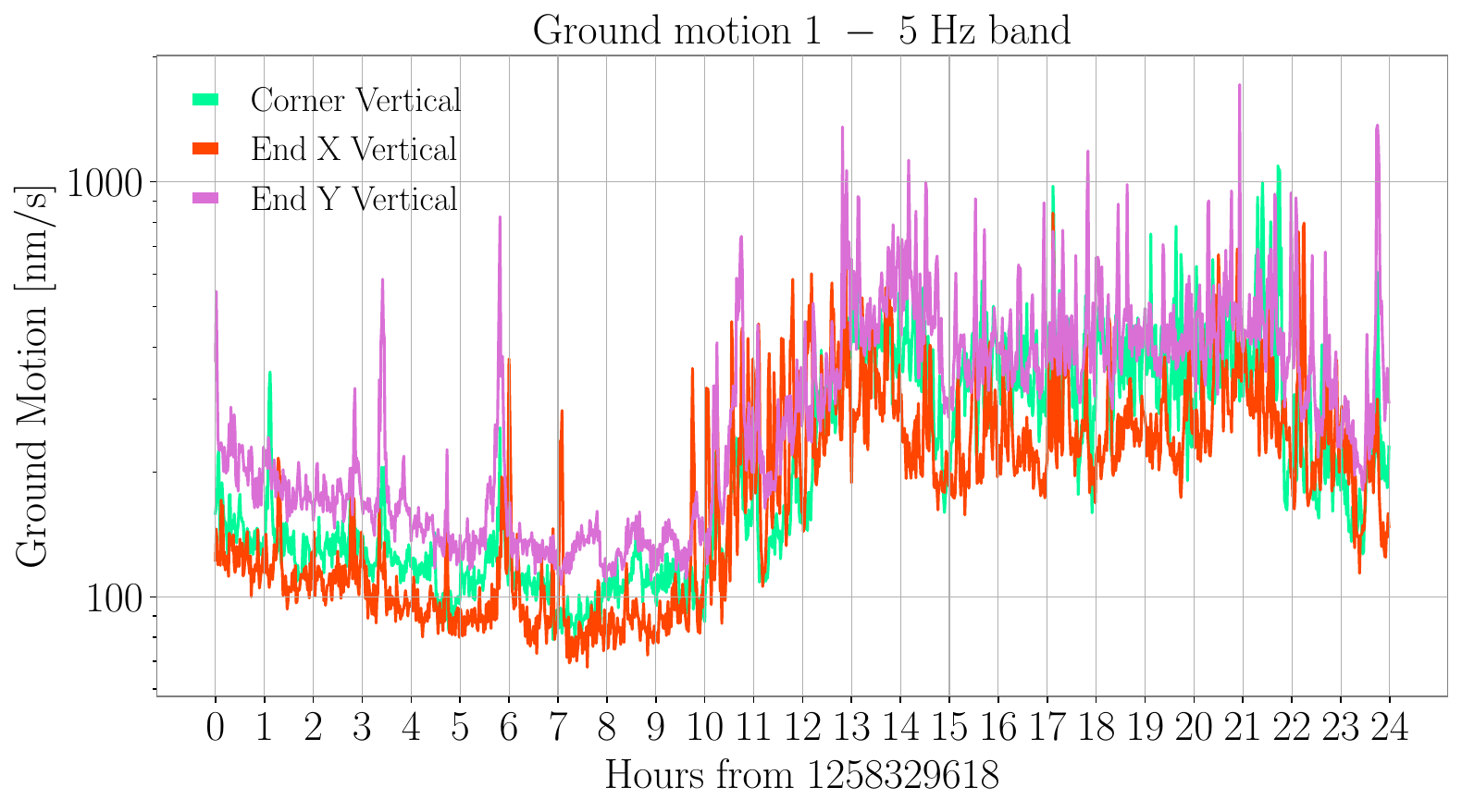}
         \label{fig:nov21gm}
    \end{subfigure}
    \hfill
    \begin{subfigure}[b]{0.48\textwidth}
        \centering
         \includegraphics[width =\textwidth]{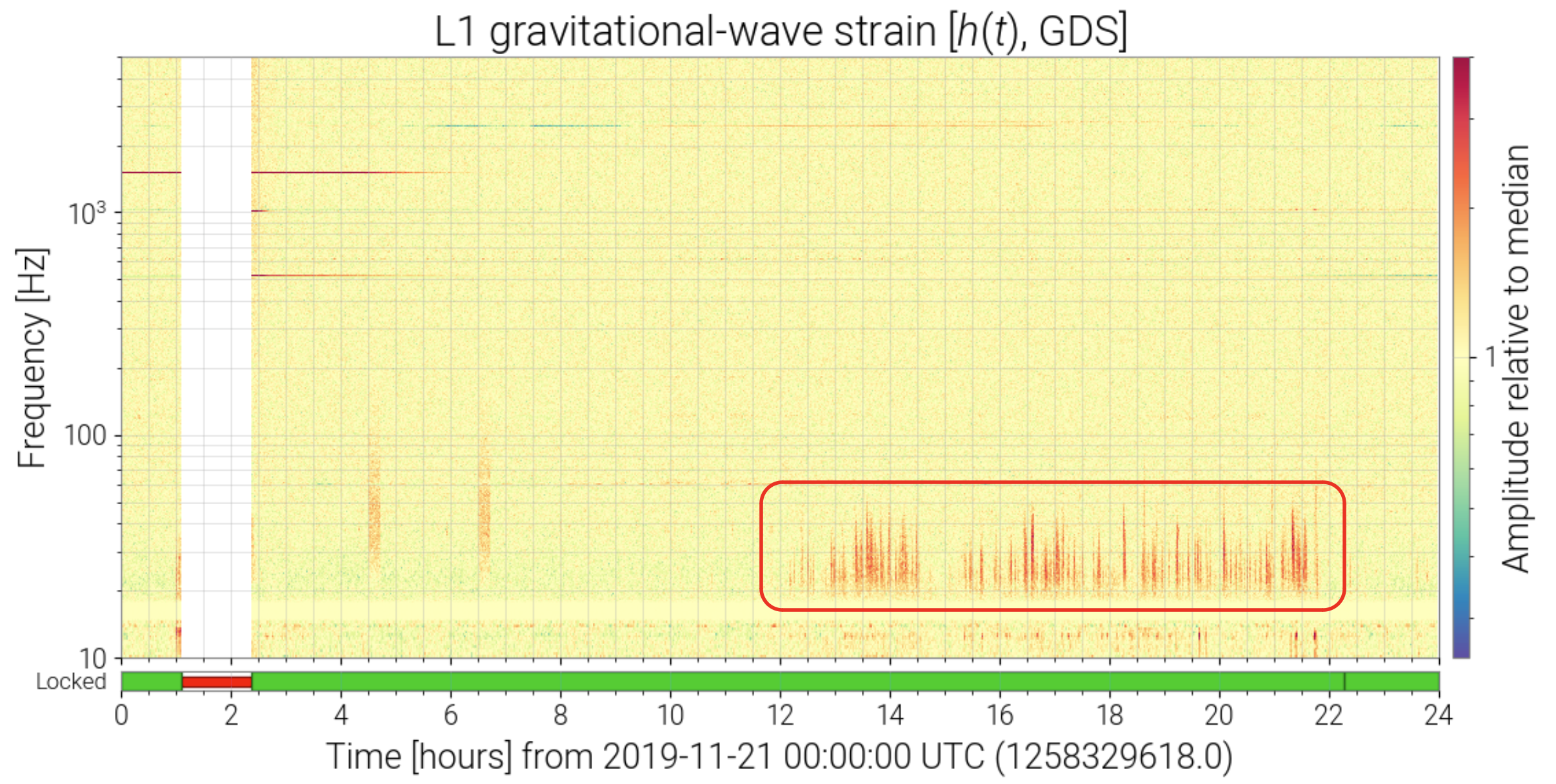}
         \label{fig:nov21strain}

    \end{subfigure}
    
    \caption{Comparison of the rise in anthropogenic motion and the noise in h(t) strain spectrogram on Nov 21 2019 after 12:00:00 UTC. \emph{Left}: Post noon UTC, the anthropogenic ground motion floor at the Corner as well as the X and Y End stations is elevated due to human activity. \emph{Right}: This increase in seismic activity leads to fast scatter or daytime noise in h(t) as shown in the spectrogram here. This plot was taken from the Detector \href{https://ldas-jobs.ligo-la.caltech.edu/~detchar/summary/day/20191121/lock/strain/}{Summary pages} \cite{duncan_macleod_2021_4975045}.}
    \label{fig:comparisongm}
    
\end{figure}

Figure \ref{fig:comparisongm} shows the noise in h(t) strain spectrogram and the $1$--$3~\mathrm{Hz}$ band ground motion for Nov 21 2019. Due to logging activities near the detector, the $1-10~\mathrm{Hz}$ seismic band was elevated for most days of this week \cite{alog:logging_nov21}. The majority of this noise is 4 Hz fast scatter.

\subsection{2 Hz Fast Scatter}
2 Hz fast scatter glitches have arches separated by $\sim$ 0.5 seconds as shown in the left plot in Figure \ref{fig:fast_population}. The 2 Hz population is observed to show a stronger correlation with the ground motion in the microseismic band along with the anthropogenic band. 
Ocean waves interacting with the ocean floor is the dominant cause of increased microseism near the detector \cite{microsiesm_paper}.
Microseismic ground motion is also seasonal and is generally  higher during the winter months. The rate of the 2 Hz population was observed to be significantly higher during Feb 2020 at LLO. This higher rate can be explained by an increased microseism and reduced rate of slow scatter during February 2020. Before the implementation of RC tracking in January 2020, slow scattering was the dominant transient noise during high microseism. Post RC tracking, the reduction in the slow scattering transients contributed to increased visibility of 2 Hz fast scatter.

\begin{figure}[ht]
\captionsetup[subfigure]{font=scriptsize,labelfont=scriptsize}
   \centering
    \begin{subfigure}[b]{0.48\textwidth}
        \centering
         \includegraphics[width= \textwidth]{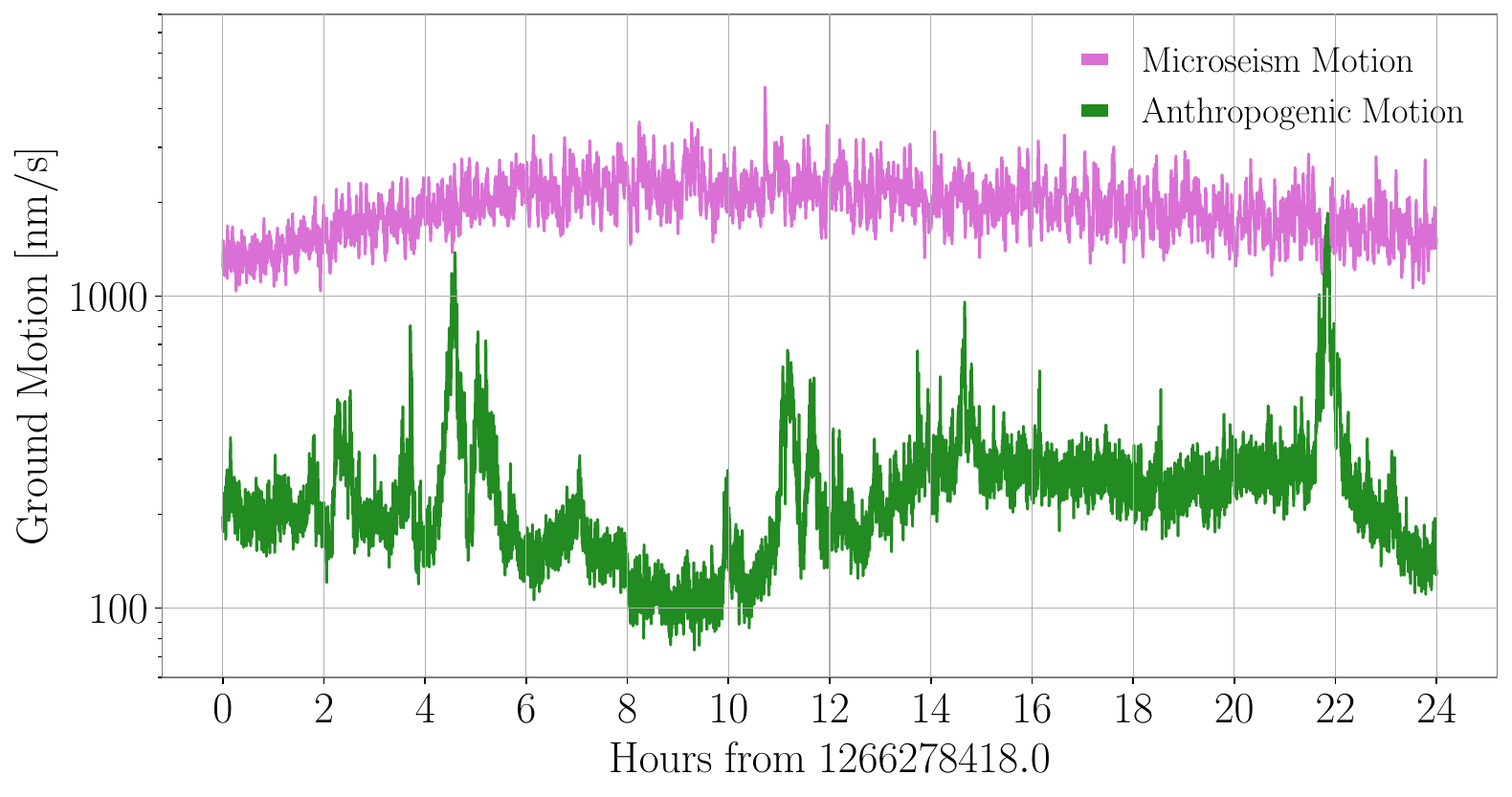}
         \label{fig:feb21gm}
    \end{subfigure}
    \hfill
    \begin{subfigure}[b]{0.48\textwidth}
        \centering
         \includegraphics[width =\textwidth]{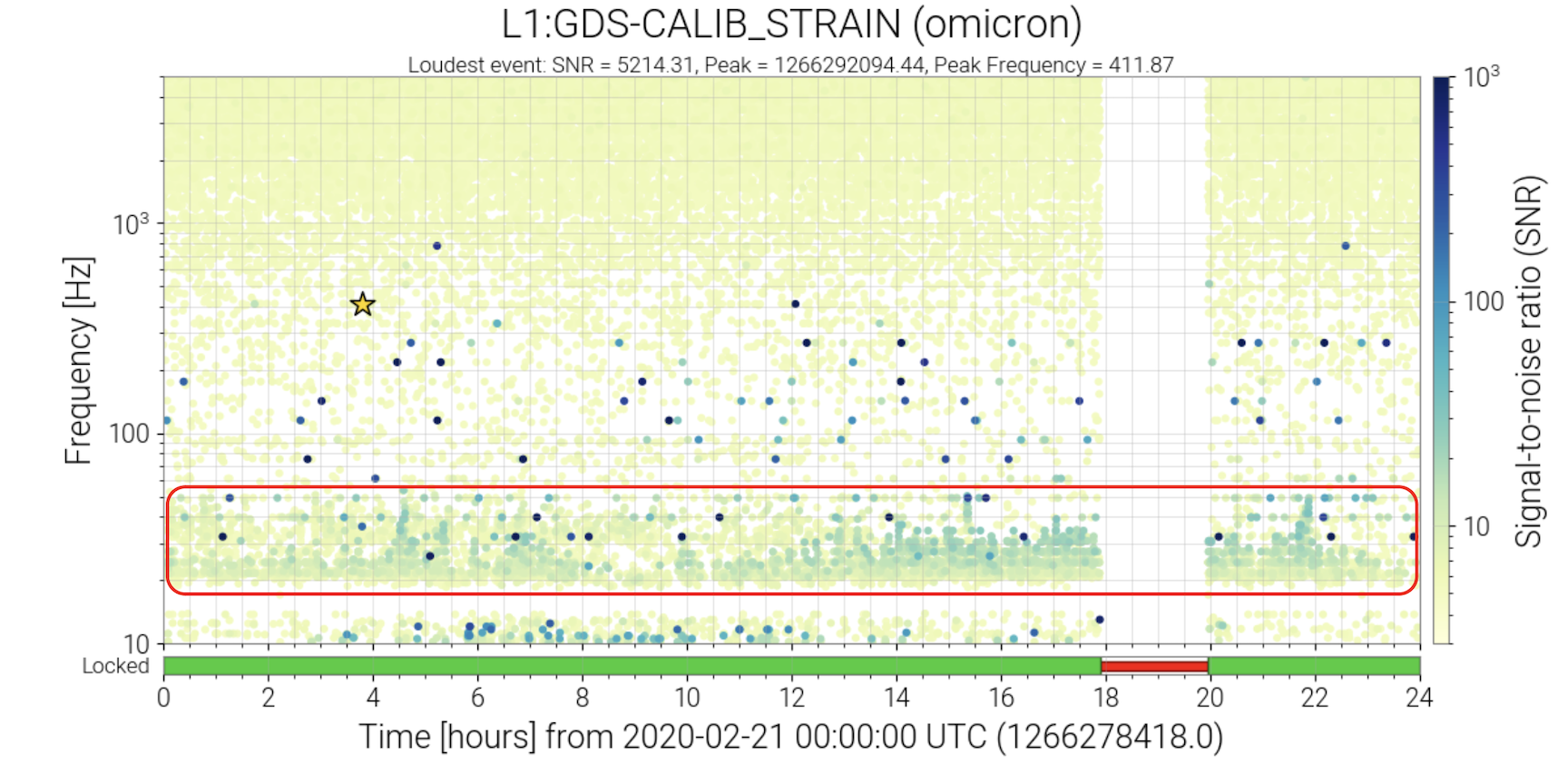}
         \label{fig:feb21glitches}
    \end{subfigure}
    \caption{\emph{Left}: The microseismic motion was very high on Feb 21 2020 and exceeded $2 \mu/s$ at times as seen in this plot. The combination of high anthropogenic and microseism motion increases the rate of fast scatter in the data. \emph{Right}: This amalgam of the two bands led to a very high number of 2 Hz Fast Scatter on this day. Most of the Omicron triggers in the red box are classified as Fast Scattering by Gravity Spy. This plot was taken from Detector \href{https://ldas-jobs.ligo-la.caltech.edu/~detchar/summary/day/20200221/lock/glitches/}{Summary pages} \cite{duncan_macleod_2021_4975045}.}
    \label{fig:comparisongm_2hz}
    
\end{figure}

In the Post O3 data we found 3.3 Hz fast scatter during trains. This is discussed in \ref{append_3.3Hz}.
Depending on the ground motion in different frequency bands, we have observed some fringe populations where the arches are separated by $\sim$ 0.75 seconds. These different populations discussed in this section suggests that the frequency region of the ground motion is intimately linked with the morphology and type of fast scatter we observe in the data \cite{dcc_fast_scatter_at_LLO}. We discuss this in more detail in Sec \ref{sec_noise_modelling}.

\section{Instrumental Correlations}\label{section_instrumental_corr}
The mitigation of transient noise can roughly be divided into three stages. The first stage is to find the environmental or instrumental conditions that correlate with the presence of noise. In the case of fast scatter, this would be increase in anthropogenic and/or microseismic ground motion. The next step is to localize the potential noise coupling site in the complex system of instrument hardware. This could be an optic moving under the influence of increased ground motion in one of the detector subsystems. The final step is to make the necessary hardware modifications at the source, to reduce or eliminate the transient noise.

The three potential sites where  fast scatter can originate from are the Y End Station, X End Station and the Corner Station where the majority of the detector optics and other hardware are located. Usually, the ground motion does not differ very much across the three stations. However, certain incidents result in higher ground motion at one station compared to others. These special conditions include passing trains which result in higher ground motion at the Y End Station at LLO due to its proximity with the train tracks, road construction work and logging if the work site is closer to one of the stations compared to others. Such conditions present an ideal opportunity to study the correlation between fast scatter and the potential coupling sites. These special conditions also help in eliminating certain locations as potential sources of noise. For example, if a period of particularly high ground motion at one of these locations does not result in the expected increase of the levels of noise, that means either the noise does not couple at this location or that the coupling is weak. 

\subsection{Corner Station Correlation with 4 Hz Fast Scattering}
For a number of days in O3 with either road construction or logging near LLO, we observed a correlation between the anthropogenic ground motion at the Corner station and changes in the differential arm length (DARM). Next we look at some of these days:

\begin{figure}[ht]
\captionsetup[subfigure]{font=scriptsize,labelfont=scriptsize}
   \centering
    \begin{subfigure}[b]{0.48\textwidth}
        \centering
         \includegraphics[width= \textwidth]{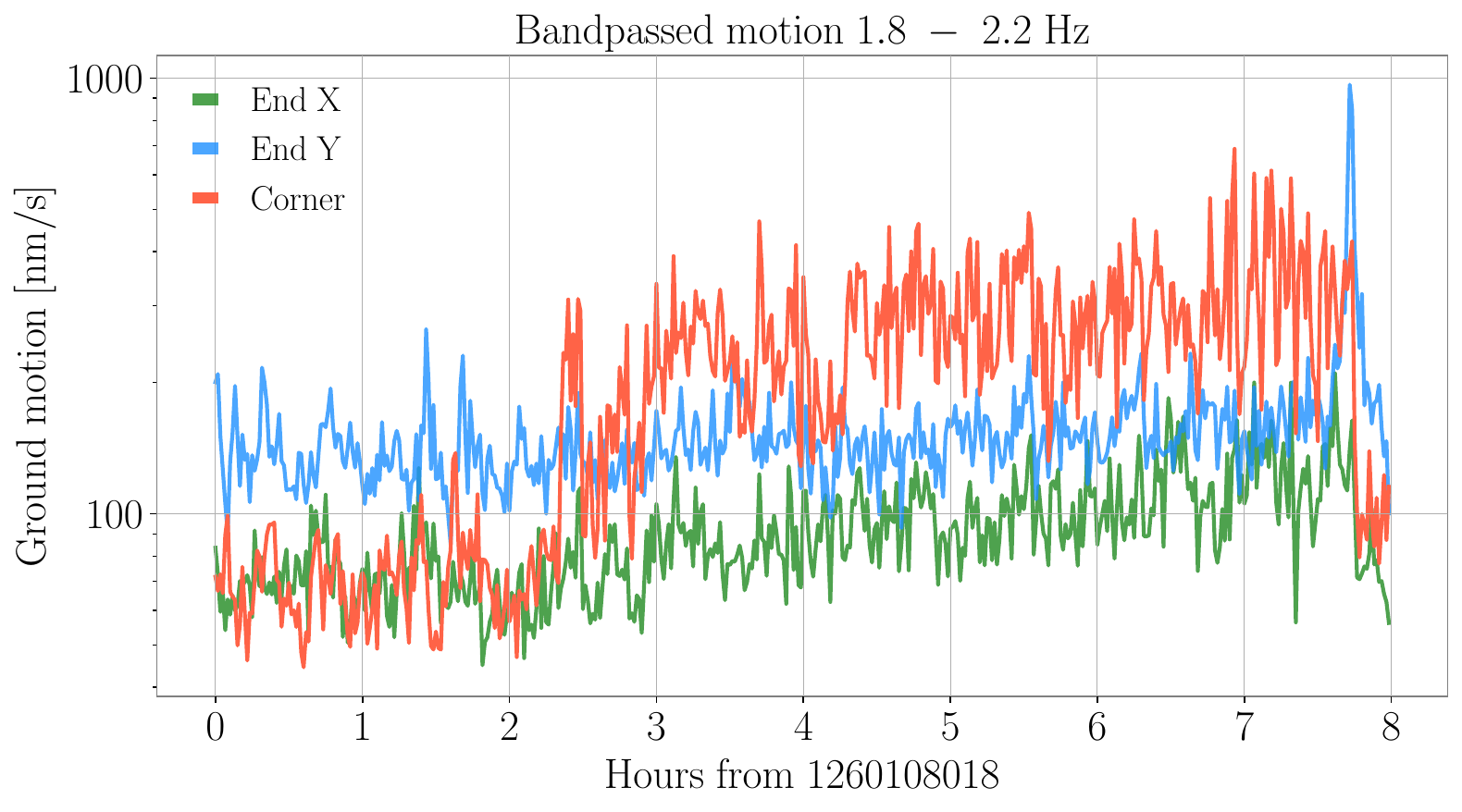}
         \label{fig:dec11_gm}
    \end{subfigure}
    \hfill
    \begin{subfigure}[b]{0.48\textwidth}
        \centering
         \includegraphics[width =\textwidth]{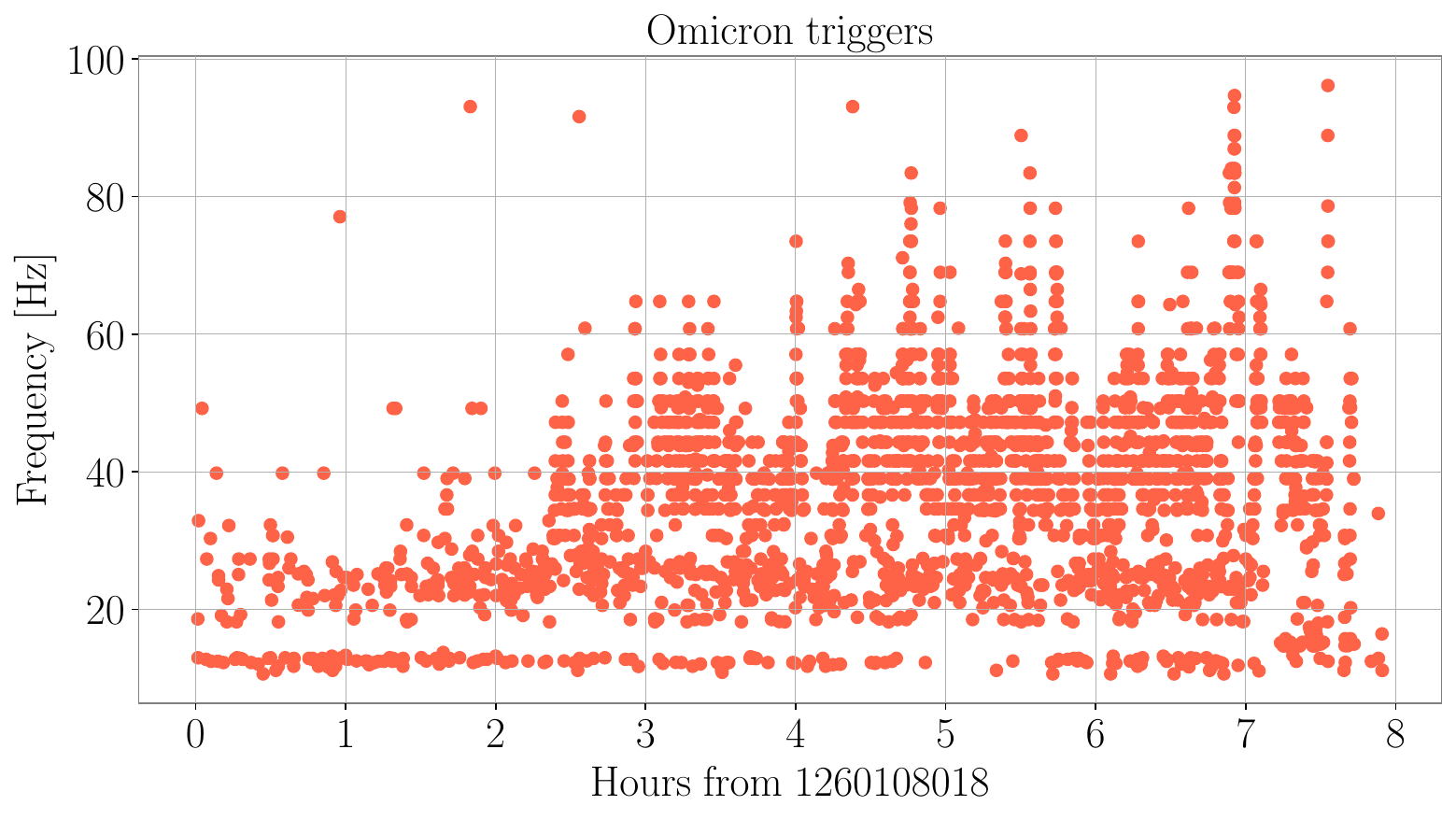}
         \label{fig:dec11glitches}    
    \end{subfigure}
    \caption{Ground motion bandpassed between $1.8$ and $2.2$ Hz at X, Y and Corner station on Dec 11 2019. The Corner station motion floor gets elevated between the 2 and 3 hour mark. In comparison, the X and Y End station ground motion amplitude does not experience such a prominent shift. \emph{Right}: Omicron triggers in the band $10-100~\mathrm{Hz}$. Between the $2$ and $3$ hour mark, there is a noticeable increase in the number of Omicron triggers which coincides with the rise in Corner station motion.}
    \label{fig:dec11_gm_Omicron}
    
\end{figure}
\subsubsection*{Dec 11 2019}
Between Dec 5 2019 and Dec 15 2019, the anthropogenic ground motion in the band $1-10~\mathrm{Hz}$ was high due to logging near the site \cite{alog:logging_dec2019}. During this period, there were several days when the anthropogenic ground motion at the Corner station was noticeably higher compared to the End stations. For these days, we observed a clear correlation between the Corner station ground motion and fast scatter noise. Figure \ref{fig:dec11_gm_Omicron} shows this correlation for Dec 11 2019. As seen in the first plot, between the $2.5$ and $8$ hour mark, the ground motion at the Corner station is elevated, compared to the first $2.5$ hours. During this period, there is a sharp increase in the amount of Omicron triggers with frequency above $\sim$ 30 Hz. There is no substantial change in the End X and End Y ground motion during this time. Such a correlation between the Corner station and the rate of fast scatter was noticed for several days in the first two weeks of Dec 2019 \cite{alog_comment:corner_correlation_dec2019}. 

\begin{figure}[ht]
\captionsetup[subfigure]{font=scriptsize,labelfont=scriptsize}
   \centering
    \begin{subfigure}[b]{0.48\textwidth}
        \centering
         \includegraphics[width= \textwidth]{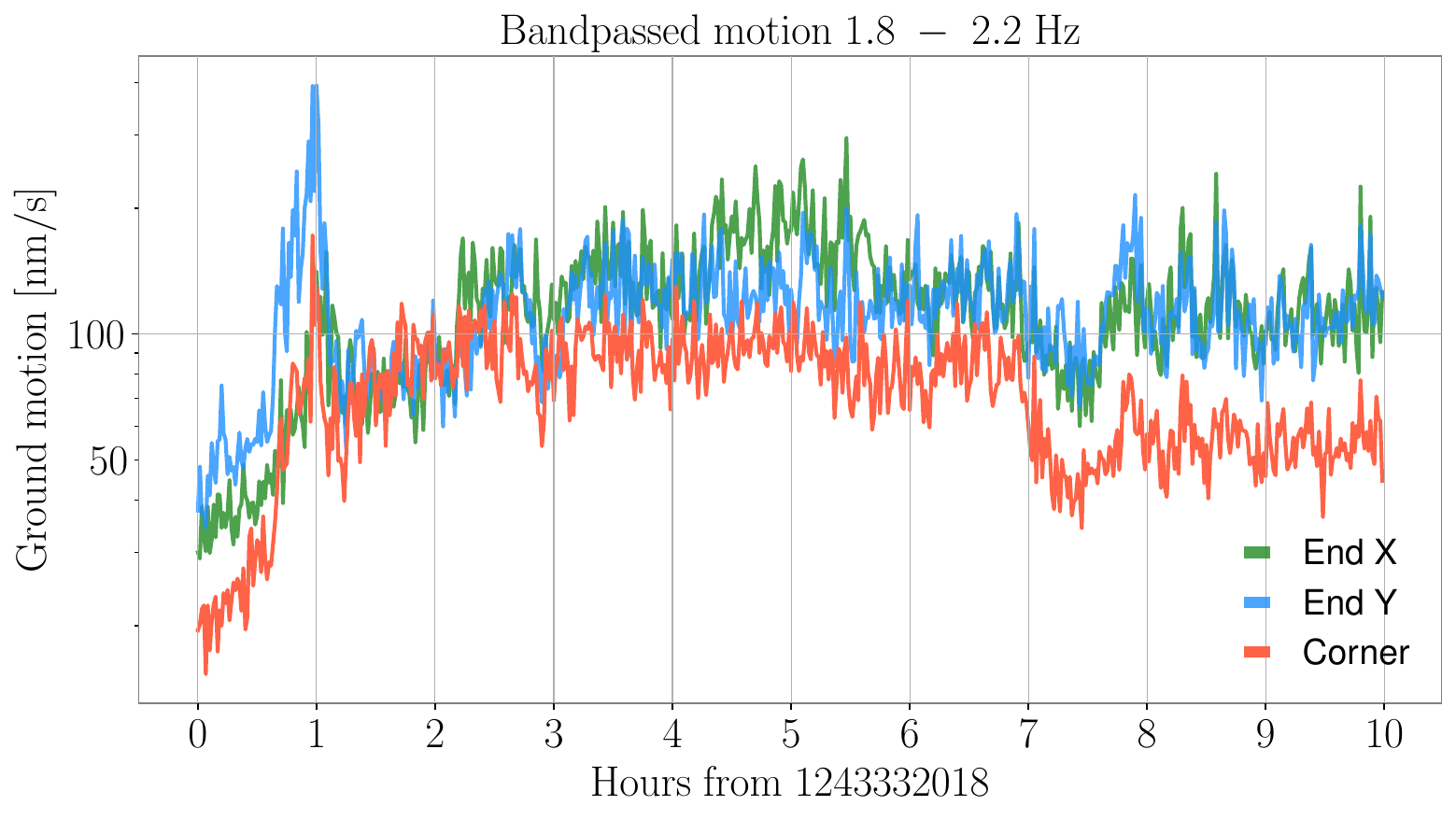}
         \label{fig:may31_gm}
    \end{subfigure}
    \hfill
    \begin{subfigure}[b]{0.48\textwidth}
        \centering
         \includegraphics[width =\textwidth]{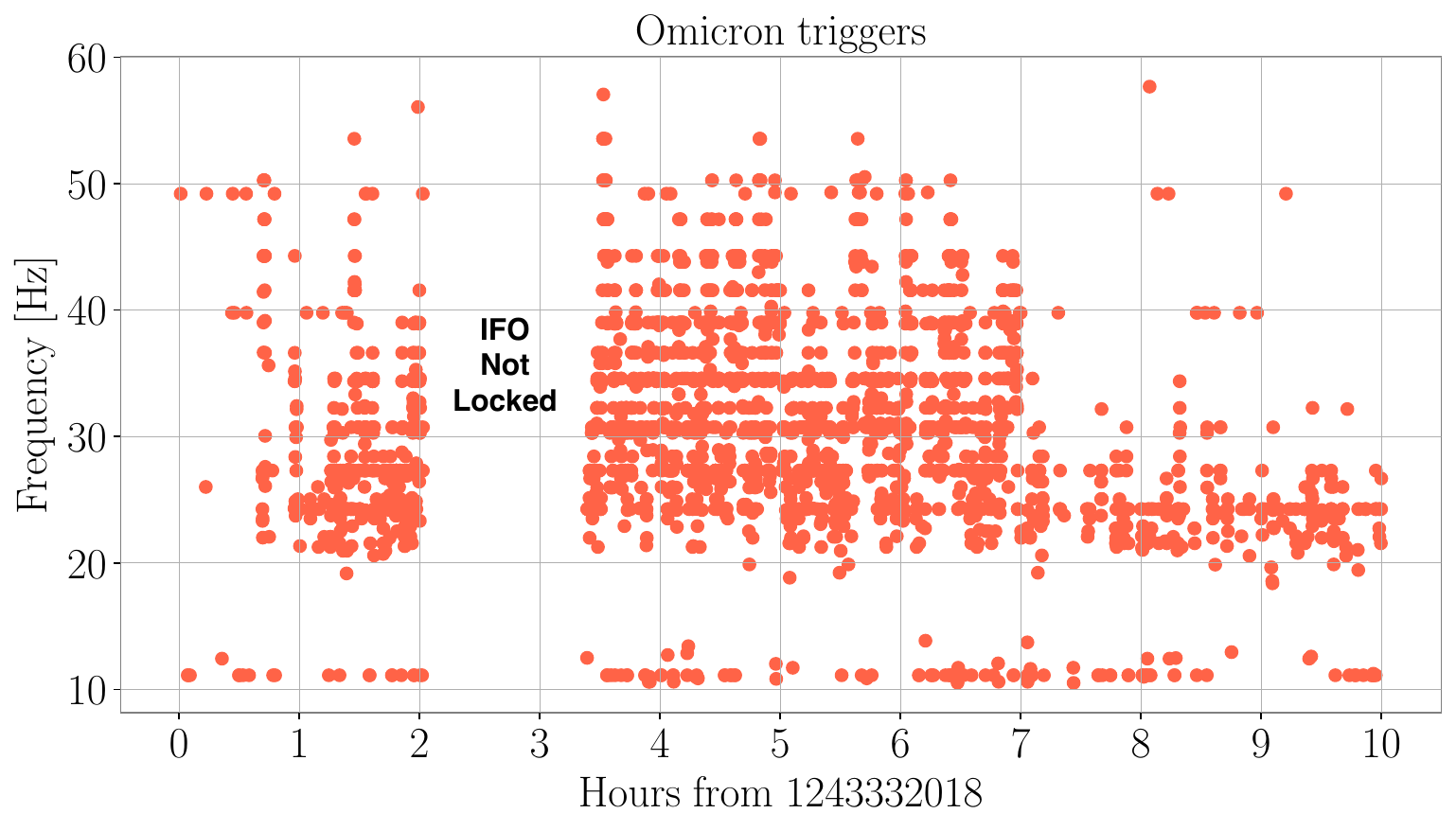}
         \label{fig:may31glitches}    
    \end{subfigure}
    \caption{\emph{Left}: Ground motion bandpassed between $1.8$ and $2.2$ Hz at X, Y and Corner station on May 31 2019. Between the $7$ and $8$ hour mark, the Corner station motion undergoes a comparatively more prominent reduction. \emph{Right}: Omicron triggers in the band $10-60~\mathrm{Hz}$. The number of triggers detected by Omicron goes down around the same time the Corner station motion starts to diminish.}
    \label{fig:may31_gm_Omicron}
    
\end{figure}
\subsubsection*{May 31 2019} This is another day when anthropogenic ground motion variation between the Corner station and End stations is more easily visible. As we can see from the left plot in Figure \ref{fig:may31_gm_Omicron}, the Corner station ground motion is relatively high between the 1 and 7 hour mark. During the same period, there is an excess of Omicron triggers in the frequency range of $20-60~\mathrm{Hz}$. As the ground motion subsides and stays low between $17$ and $20$ UTC, there are visibly less transients. 
The End stations ground motion however, does not register any substantial change and thus cannot explain the reduction in noise.

\begin{table}[]
    \centering
        \input{table_fs_corr}
    \caption{Spearman correlation between ground motion captured at different stations and Omicron transient rate for three days in O3 with high ground motion in $1.8-2.2~\mathrm{Hz}$ band. The seismic motion and the Omicron transients for Dec 11 2019 and May 31 2019 are shown in Figure \ref{fig:dec11_gm_Omicron} and Figure \ref{fig:may31_gm_Omicron}. As shown in these figures and this table, for all these three days, the Corner station motion shows the best correlation with the transient rate. }
    \label{tab:fs_corr}
\end{table}


In Table \ref{tab:fs_corr}, we show the Spearman correlation between the anthropogenic ground motion at the Corner and End stations and fast scatter noise for three days in O3 \cite{boslaugh_watters_2009}. In the next section we look at how the $2$ Hz motion can create $4$ Hz fast scatter in h(t). In Section \ref{section_acb}, we look at the potential suspects in the Corner station responsible for this noise.

\section{Noise Modelling}\label{sec_noise_modelling}
Using equations \ref{eq:1} and \ref{eq:2}, we can model fast scatter noise. As mentioned earlier, the 2 Hz Fast scatter is correlated with an increase in microseismic band whereas the rate of 4 Hz increases with an increase in seismic noise in the anthropogenic band. We want to understand if the combination of ground motion in different bands can explain the different populations of fast scattering in the data. The following equation represents our noise model:

\begin{equation}
    h(t) = \mathrm{quiet\_data} + \mathrm{phase\_noise}(a_1*v_{\mathrm{1}}(t) + a_2*v_{\mathrm{2}}(t))
\end{equation}

The $v_{1}$ and $v_{2}$ represents \textbf{anthropogenic} and \textbf{microseismic} velocity respectively, while $a_{1}$ and $a_{2}$ are amplitude knobs to modulate the motion. Using this model, depending on the amount of microseismic motion added to 2 Hz anthropogenic motion, we can generate both the 2 Hz and the 4 Hz fast scatter noise \cite{dcc_fast_scatter_at_LLO, dcc_LVK_mar_2023_scatter}. 
If the injected microseism motion at 0.15 Hz is less than one-half of the anthropogenic motion at 2 Hz, then the combination of 2 Hz and 0.15 Hz motion shows up as 4 Hz fast scatter. If we then increase the amount of injected 0.15 Hz motion, we obtain 2 Hz scatter. Essentially:


\begin{equation}
    \frac{\mathrm{a_1}}{\mathrm{a_2}} \gtrapprox \mathrm{threshold\rightarrow 4 \ Hz \ noise}
\end{equation}

\begin{equation}
    \frac{\mathrm{a_1}}{\mathrm{a_2}} \lessapprox \mathrm{threshold\rightarrow 2 \ Hz \ noise}
\end{equation}

This change from 4 Hz to 2 Hz fast scatter noise is gradual with the increase in microseismic motion.
\begin{figure}[ht]
\captionsetup[subfigure]{font=scriptsize,labelfont=scriptsize}
   \centering
    \begin{subfigure}[b]{0.45\textwidth}
        \centering
         \includegraphics[width= 1\textwidth]{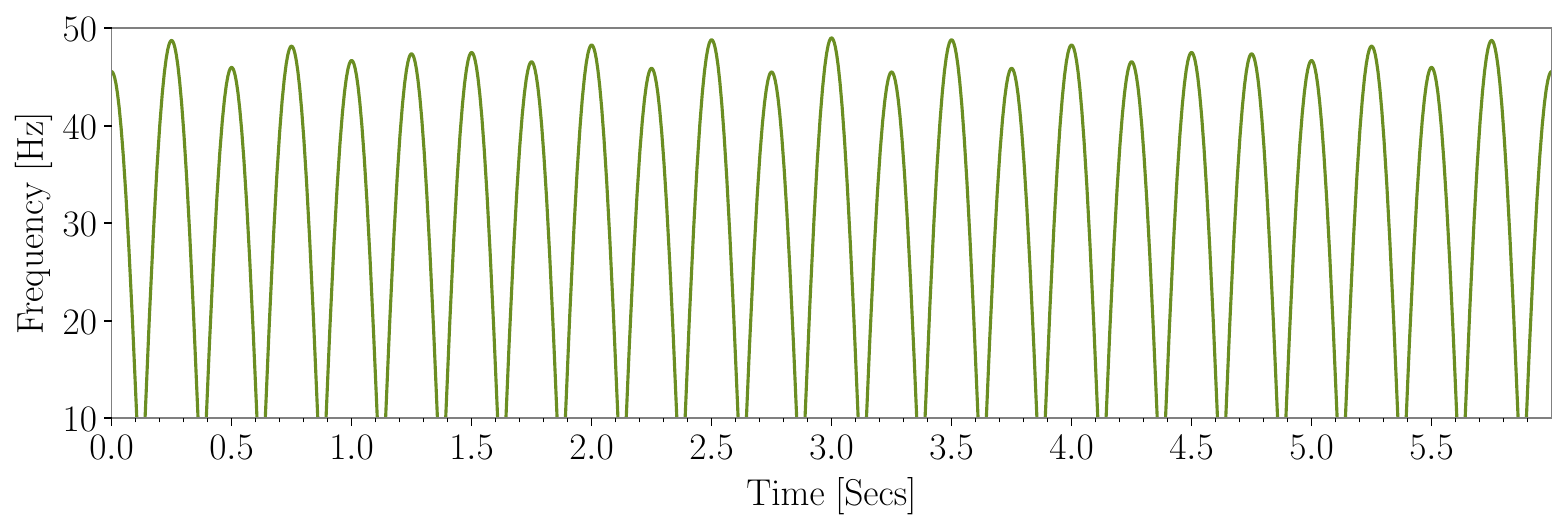}
         \label{fig:4hz_fs}
    \end{subfigure}
    \hfill
    \begin{subfigure}[b]{0.45\textwidth}
        \centering
         \includegraphics[width =\textwidth]{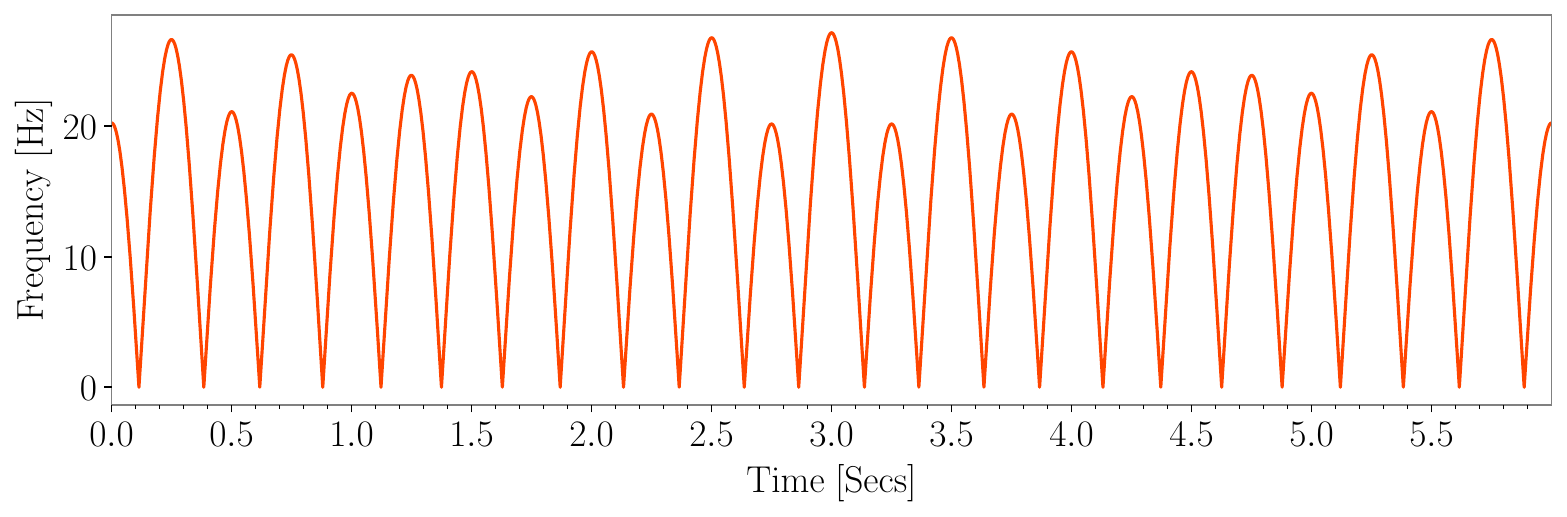}
         \label{fig:2hz_ts}    
    \end{subfigure}

    \begin{subfigure}[b]{0.45\textwidth}
        \centering
         \includegraphics[width =\textwidth]{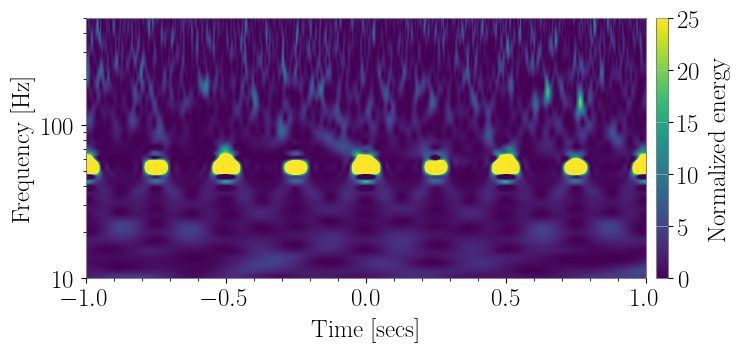}
         \label{fig:4hz_phase}    
    \end{subfigure}
    \hfill
    \begin{subfigure}[b]{0.45\textwidth}
        \centering
         \includegraphics[width =\textwidth]{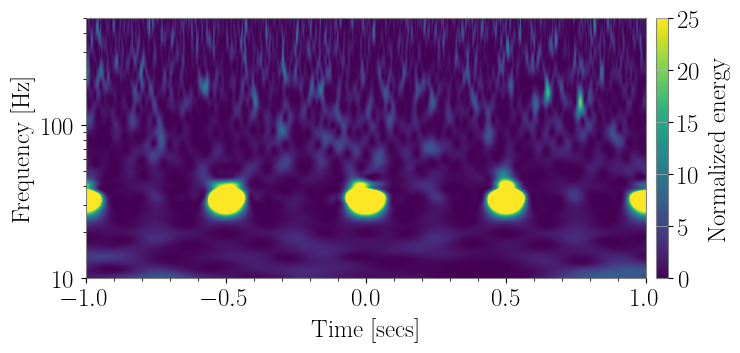}
         \label{fig:2hz_phase}    
    \end{subfigure}
     \caption{Fringe frequency and spectrogram of the phase noise modelled by injecting anthropogenic and microseismic motion and $2$ Hz and $0.15$ Hz respectively \emph{Left}: Here the ratio of the amplitude for the $2$ Hz and $0.15$ Hz motion is 2 and the peaks in the fringe frequency motion and the arches in the spectrogram are separated by $\sim$ 0.25 seconds. So when the anthropogenic noise dominates, $2$ Hz motion show up as $4$ Hz noise. \emph{Right}: Now we increase the amount of microseism motion so the ratio of $2$ Hz and $0.15$ Hz motion injected is  $\frac{1}{2}$. The dominant peaks in the fringe frequency motion and the spectrogram arches are now separated by $0.5$ seconds. For increased microseism in the data, the $2$ Hz motion shows up as $2$ Hz noise. }
    \label{fig:2hz4hz_ts}
    
\end{figure}
\subsection{High anthropogenic, low microseism}
We first look at the case where $\frac{a_{1}}{a_{2}} > 2$. We add half as much microseismic motion at $0.15$ Hz to anthropogenic motion at 2 Hz. The resultant fringe frequency motion and the phase noise is at 4 Hz. This is shown in the plots on the left in Figure \ref{fig:2hz4hz_ts}. This is also what we observe in O3 data,  days with high anthropogenic but low microseism motion are dominated by $4$ Hz fast scatter noise.

\subsection{High anthropogenic, high microseism}
Next, we increase the amount of relative microseism motion in our model. The plots on the right in Figure \ref{fig:2hz4hz_ts} show the fringe frequency and the phase noise when for $\frac{a_{1}}{a_{2}} = 1/2$. Once again, we use $2$ Hz and $0.15$ Hz for $v_{1}$ and $v_{2}$ respectively. In O3, 2 Hz fast scatter was prevalent in Feb 2020 when we had increased microseismic activity. As we increase the microseism and the ratio $\frac{a_{1}}{a_{2}}$ changes from 2 to $\frac{1}{2}$, we can see the transition from 4 Hz noise to 2 Hz in the spectrograms.


\begin{figure}[ht]
\captionsetup[subfigure]{font=scriptsize,labelfont=scriptsize}
   \centering
    \begin{subfigure}[b]{0.48\textwidth}
        \centering
         \includegraphics[width= 0.8\textwidth,height=1\textwidth]{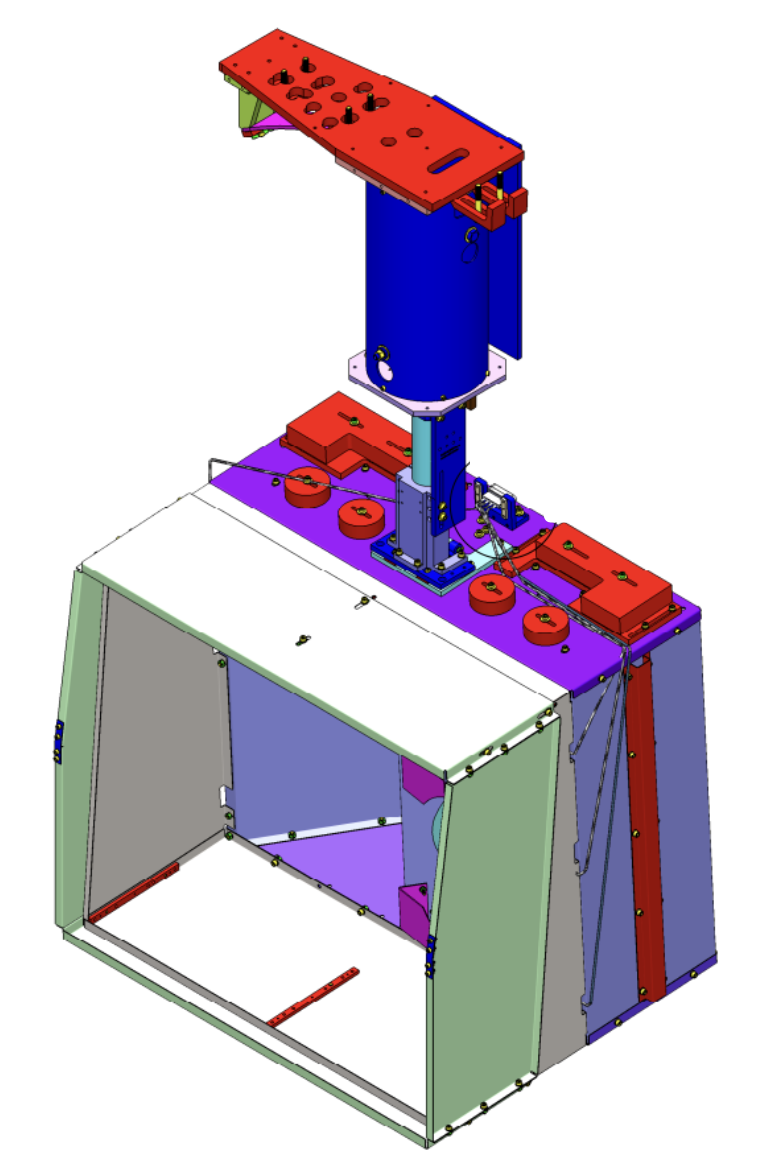}
         \label{fig:acb_schematic}
    \end{subfigure}
    \hfill
    \begin{subfigure}[b]{0.48\textwidth}
        \centering
         \includegraphics[width =\textwidth]{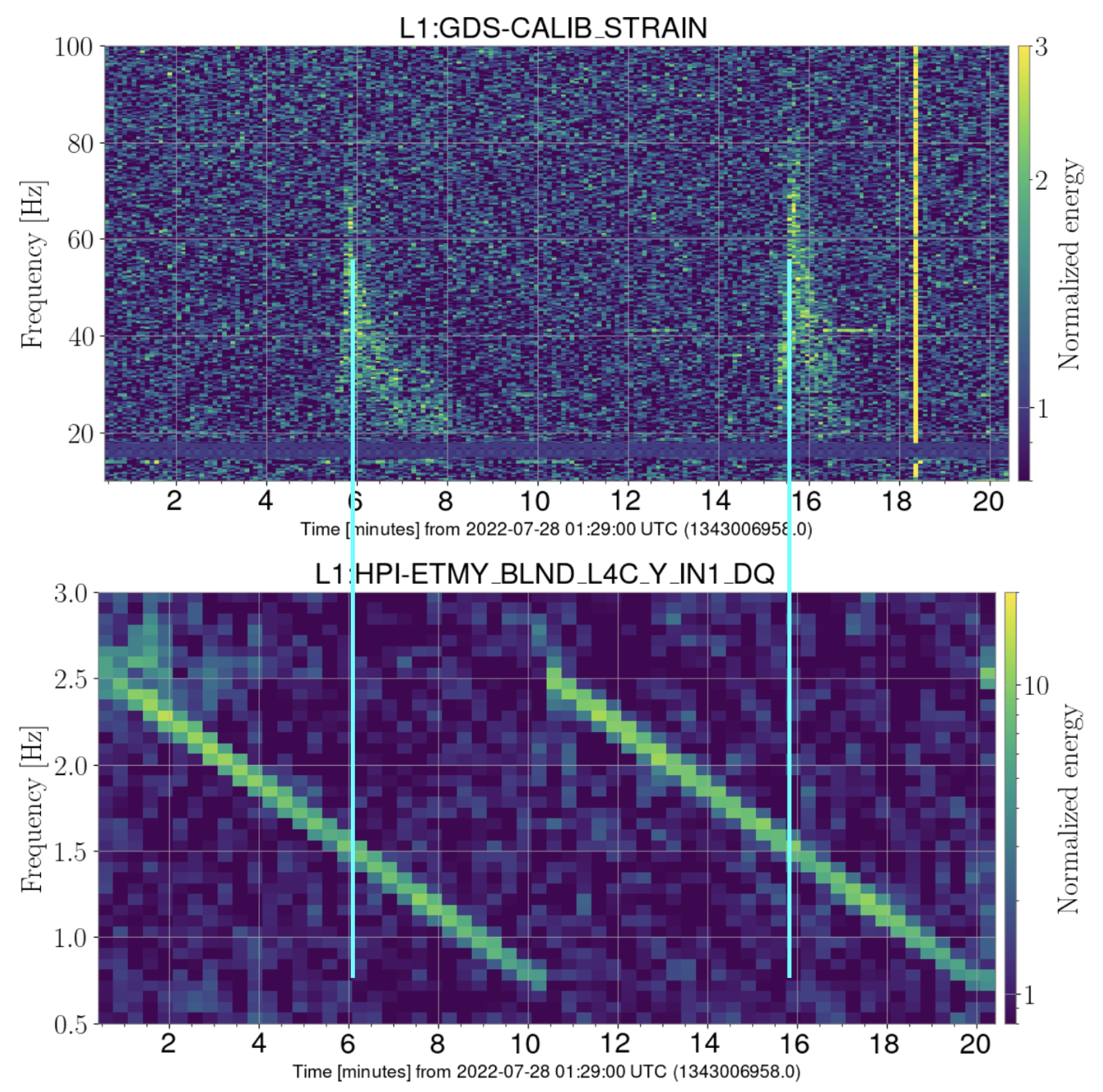}
         \label{fig:etmy_acb_jul22}    
    \end{subfigure}
    \caption{\emph{Left}: Schematic of the Arm Cavity Baffle (ACB). The ACB is a wide rectangular baffle installed in front of both the End Test Masses (ETM's) and Input Test Masses (ITM's). \emph{Right}: PEM injection sweeps that primarily excited the ACB revealed the presence of high Q resonances close to $1.6$ Hz at the Corner and Y End station ACBs. The bottom plot is a normalized spectrogram of the input motion to the ACB; as the injected motion crosses $\sim$ 1.6 Hz, fringing noise is observed in the h(t) spectrogram shown in the top plot. The large duration in h(t) is an indication of a high Q factor decay time. The test is repeated twice to rule out any accidental coincidences. \cite{alog:etmy_acb_noise}.}
    \label{fig:acb_etmy_inj}

\end{figure}
Figure \ref{fig:2hz4hz_ts} demonstrates how the ratio of different microseismic and anthropogenic amplitudes vary the type of fast scattering we observe. Not as prevalent but the O3 Fast Scatter data does contain transients where the arch separation is $\sim$ 0.75 secs. The model discussed in this section can also simulate this population.

\section{Baffle Resonances}\label{section_acb}
Scattered light baffles are installed at multiple locations in the detector to prevent scattered light from recombining into the main beam and producing scattering noise \cite{dcc_stray_light_design, dcc_stray_light_baffles_design}. However, if these structures are not properly damped, they can amplify the input motion at their mechanical resonant mode frequencies. The greater motion can cause scattering noise at frequencies that are in the sensitive band of the detector, even if the mechanical resonance frequencies are below the sensitive band.   


\subsection{Noise from cryo-manifold baffles at LHO and LLO}
Before each LIGO observing run, there is a formal program of noise injections to determine the sensitivity of the detector to the environment (Physical Environment Monitoring (PEM) injections: \cite{AdvLIGO:2021oxw, effler2015environmental}). Just before the O3 observation run, these injection showed that there were vibration sensitivities at the Y-End stations at each site that were likely due to scattering noise sources. The vibrating surfaces producing the noise at LHO were identified at the end of the O3 run \cite{alog:cryobaffle_LHO_EY} using a movie technique that had previously helped identify reaction masses as a source of scattering noise \cite{alog:AERMscatter_LHO_2020, Soni_2021}. Frames were analyzed from movies of the inside of the vacuum enclosure when scattering noise was present (either from vibration injections or, later, from vibrations produced by a wind storm). The analyses showed that light from the cryo-manifold baffle was modulated with a long decay and a fundamental frequency similar to those of the scattered light noise, suggesting that this baffle was the source of the noise. 

The cryo-manifold baffles (CB), are present in front of each of the four test masses. Their purpose is to shield reflective surfaces from light scattered from the test mass, in particular, shielding a beam-tube reduction flange at the end of a beam-tube manifold, and a cryogenic pump. After the O3 observing run, in the spring of 2020, shaker injections near the locations of 3/4 of the cryo-manifold baffles at each detector site produced scattering noise in the gravitational wave channel with a fundamental frequency of about $4$ Hz \cite{dcc_cryo_design, alog:cryobaffle_LHO, alog:cryobaffle_LHO_EY, alog:cryobaffle_EY, alog:cryobaffle_ITMY, alog:cryobaffle_ETMX}. The $4$ Hz mechanical resonances of the cryo-manifold baffle, which amplified the motion of the vacuum enclosure, were damped at three of the four baffle locations at both sites using Viton mechanical dampers. Further excitations suggested that the damping had reduced the velocity so that the scattered light noise that they produced did not reach into the sensitive frequency band of the interferometer \cite{ AdvLIGO:2021oxw,alog:cryobaffle_damped, alog:cryobaffle_damped_LHO}. 

While the CBs were the dominant source of scattering noise at LHO, a different source appears to have dominated at LLO (which turned out to be the arm cavity baffles). The scattering noise during trains and construction work actually got worse after the incursions during which the baffles were damped \cite{alog:3.3hz_scatter, alog:trains_june_2022}. 

One of the four CBs at LHO was not damped because of schedule limitations and because it produced the least noise of the four CBs.  It began producing scattering noise in the lead up to the O4 observation run, when the laser power in the arms was increased, and possibly because of relatively increased scattering with higher power \cite{alog:powerdrop_LHO_2023}. The undamped CB at LLO also began to produce increased noise during O4 and both cryobaffles should be damped in the next entry into the vacuum chambers. 

\subsection{Arm Cavity Baffle resonances at LLO}
Arm cavity baffles (ACB's)\cite{dcc_ACB_design} are attached to Stage 0 of the Hydraulic External Pre Isolator (HEPI) at ITMX, ITMY, ETMX and ETMY; the HEPI is the first stage of isolation for the test masses, from which the seismic isolation tables are also suspended, and also isolates Stage 0 from the ground motion \cite{Wen:2013boa}. The ACB mechanical resonances are nominally damped with eddy current magnets. These baffles are used to catch the wide angle scatter from the nearby test mass and any narrow angle scatter from the far test mass (4 km away). The left plot in Figure \ref{fig:acb_etmy_inj} shows the schematic for the ACB.
PEM injections during the summer of 2022 revealed the presence of noise coupling close to 1.6 Hz in the arm cavity baffles in the Corner and Y End station \cite{alog:etmy_acb_noise}. The right plot in Figure \ref{fig:acb_etmy_inj} shows the h(t) response for sweep injections at End Y ACB. As the sweep crosses $\sim$ 1.6 Hz, fringing noise appears in h(t) in the frequency band $20-100~\mathrm{Hz}$ . Fringing noise with comparatively lower amplitude was also observed during ACB injection tests at ITMY and ITMX \cite{alog:etmy_acb_noise}. The baffle is not instrumented but the h(t) response indicates a large amplification factor at this frequency and a long decay time, both consistent with a ``hidden" high Q resonance.

\begin{figure}[ht]
\captionsetup[subfigure]{font=scriptsize,labelfont=scriptsize}
   \centering
    \begin{subfigure}[b]{0.45\textwidth}
        \centering
         \includegraphics[width= 1\textwidth]{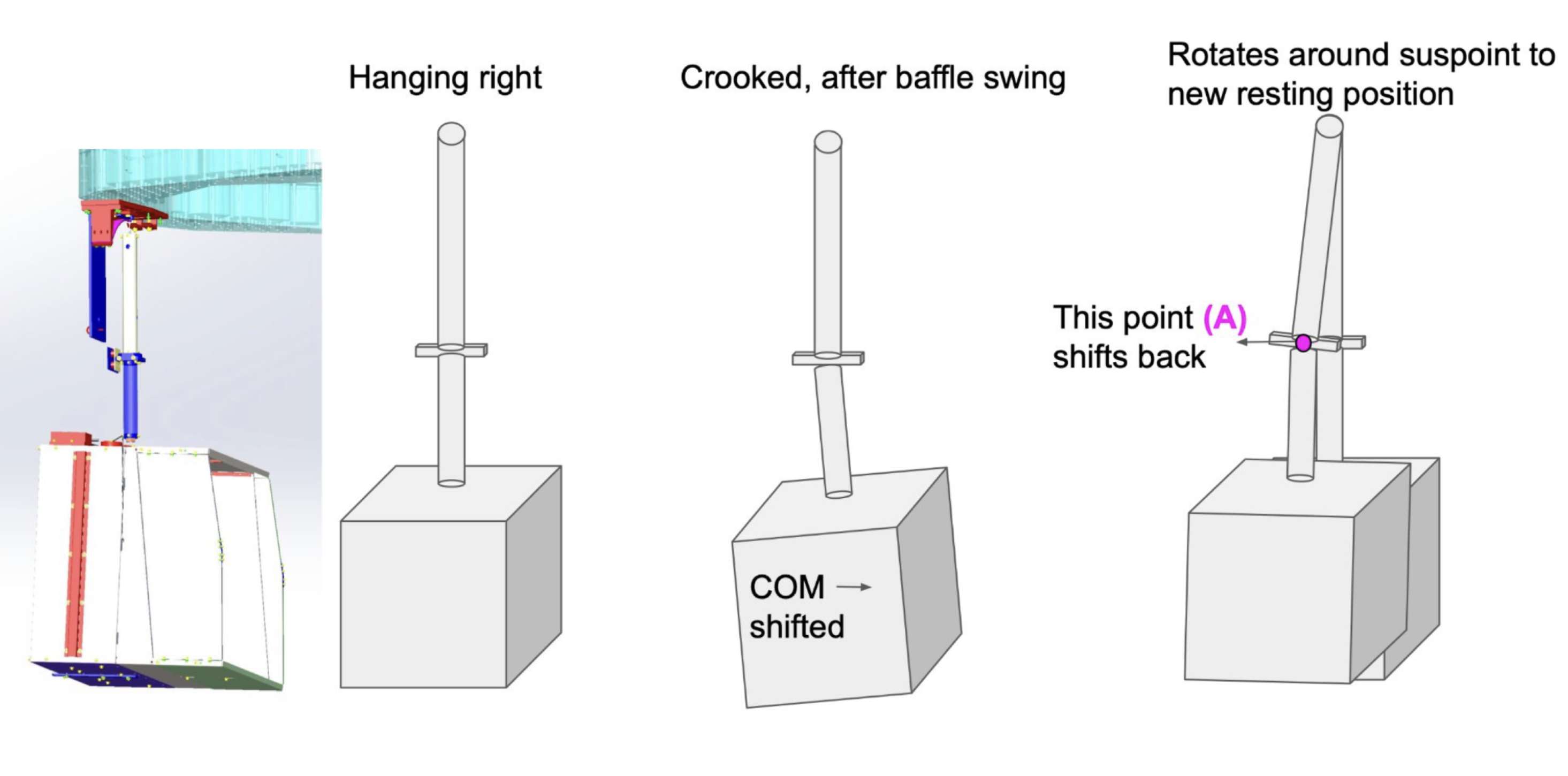}
         \label{fig:acb_position}
    \end{subfigure}
    \hfill
    \begin{subfigure}[b]{0.45\textwidth}
        \centering
         \includegraphics[width =\textwidth]{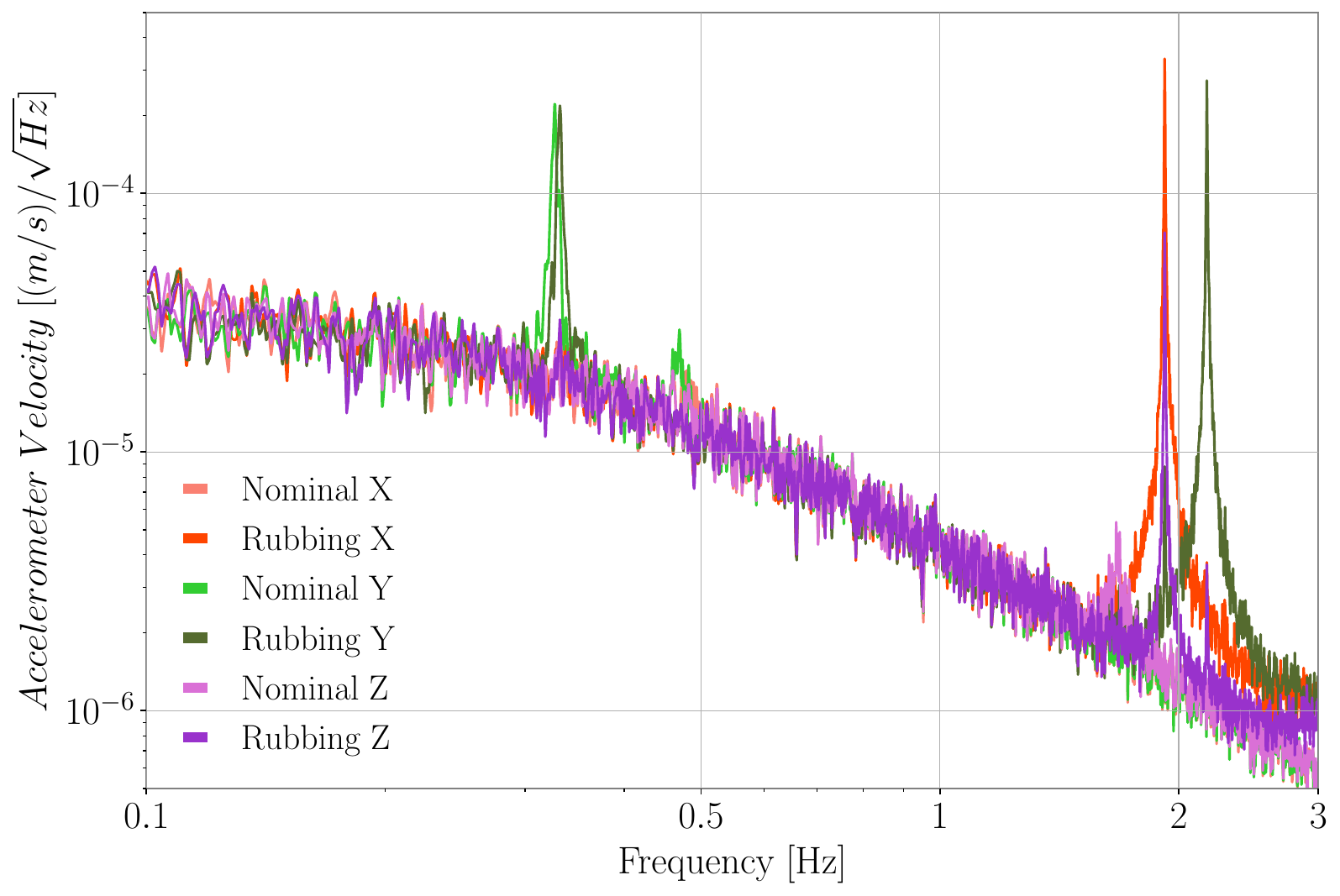}
         \label{fig:acb_rubbing}    
    \end{subfigure}
    \caption{\emph{Left}: The baffle is attached to the HEPI via 2 tubes, the lower tube (blue) and upper tube (white). The gap between the lower tube and the connector with upper tube can create a shift in the position of the baffle \cite{dcc_acb_hysteresis}. \emph{Right}: The plot shows two sets of accelerometer spectra, calibrated to velocity in m/s, for two separate conditions; each set is 3 degrees of freedom: X, Y, and Z, in the interferometer basis. One condition is the nominal state, the other is the ``rubbing", i.e. the mechanical failure that was observed and corrected. When this suspended baffle is effectively shorted through ``rubbing", high Q resonances develop as seen in the lighter color traces around 2 Hz, and furthermore affect the horizontal degrees of freedom which more easily create scattered light noise for the gravitational-wave signal. In the nominal state, the only resonance present is a very low Q, vertical direction resonance near 1.6 Hz, which does not present enough motion amplification to cause noise in the detection band. The 0.3 Hz resonance is unchanged, but it was also measured to be low enough Q to not create noise \cite{alog:acb_res_shift, alog:acb_ex_hysteresis_alog}.}

    \label{fig:acb_hysteresis_rubbing}
    
\end{figure}

In the absence of high microseism, motion at $1.6$ Hz would show up as scatter noise at $3.2$ Hz. The noise induced by trains after the O3 observing period, which appears as scatter arches separated by $\sim$ $0.3$ seconds ($3.3$ Hz), can be explained by these ACB resonances at $\sim$ 1.6 Hz. During O3 however, the fast scatter due to trains and other anthropogenic sources was at 4 Hz and 2 Hz. An interesting detail about the ACB resonant frequency is that it is very sensitive to changes in the physical configuration of the baffle. Any small changes to its state can result in a shift in the resonant frequencies. This will then change the frequency of the fringes that appear in h(t). In the absence of high microseism, ACB resonant motion at $\sim$ $2$ Hz, will create the $4$ Hz fast scatter common during O3. Addition of microseismic motion to that will give rise to the $2$ Hz fast scatter common during February 2020 in O3 \cite{dcc_fast_scatter_at_LLO}. 

The mechanism of the high Q resonance creation and the impact on the resonant frequency is shown in Figure \ref{fig:acb_hysteresis_rubbing}. The baffle is suspended from two sequential cylinders, and in between there is an attachment for the damping eddy current magnets. The baffle needs to be swung upwards, away from the test mass, to allow for test mass access during in-vacuum installation periods. Every time the baffle is swung in and out of place, the two cylinders have a chance of shifting a couple mm at the midpoint, effectively causing the baffle to shift its center of mass, swing forward and ``rub" on adjacent hardware, effectively removing the benefit of its suspension. Furthermore, because the ``rub" points are close to the eddy magnet location, their effect is bypassed since the motion induced is around this point.  In the rightmost panel of Figure \ref{fig:acb_hysteresis_rubbing} we show the spectra of 3 accelerometers (X, Y, and Z) temporarily attached to the bottom of the baffle to study this effect. The lighter, thicker traces show a free baffle, which has a resonance at 0.24 Hz and one at 1.6 Hz, but fairly low Q. The thinner, darker traces show the spectra when the baffle is ``rubbing". This can produce more than one high-Q resonance, in more than one degree of freedom, and the resonance frequency is not predictable but depends on the un-reproducible amount of ``rubbing". As such, after each vacuum incursion, the relevant ACB would have shifted its resonant frequencies and Q factors. The presence of multiple frequencies at multiple locations made this noise mechanism particularly difficult to identify. 

During the Fall of 2022, the ACB resonances found at the Corner and End Y stations were mechanically fixed by the commissioners at LLO \cite{alog:acb_fix}. The mitigation was to mechanically recenter the cylinders, allowing the baffle to hang freely and the eddy current magnets to damp the nominal resonances.

\begin{figure}
    \centering
    \includegraphics[width=0.7\textwidth]{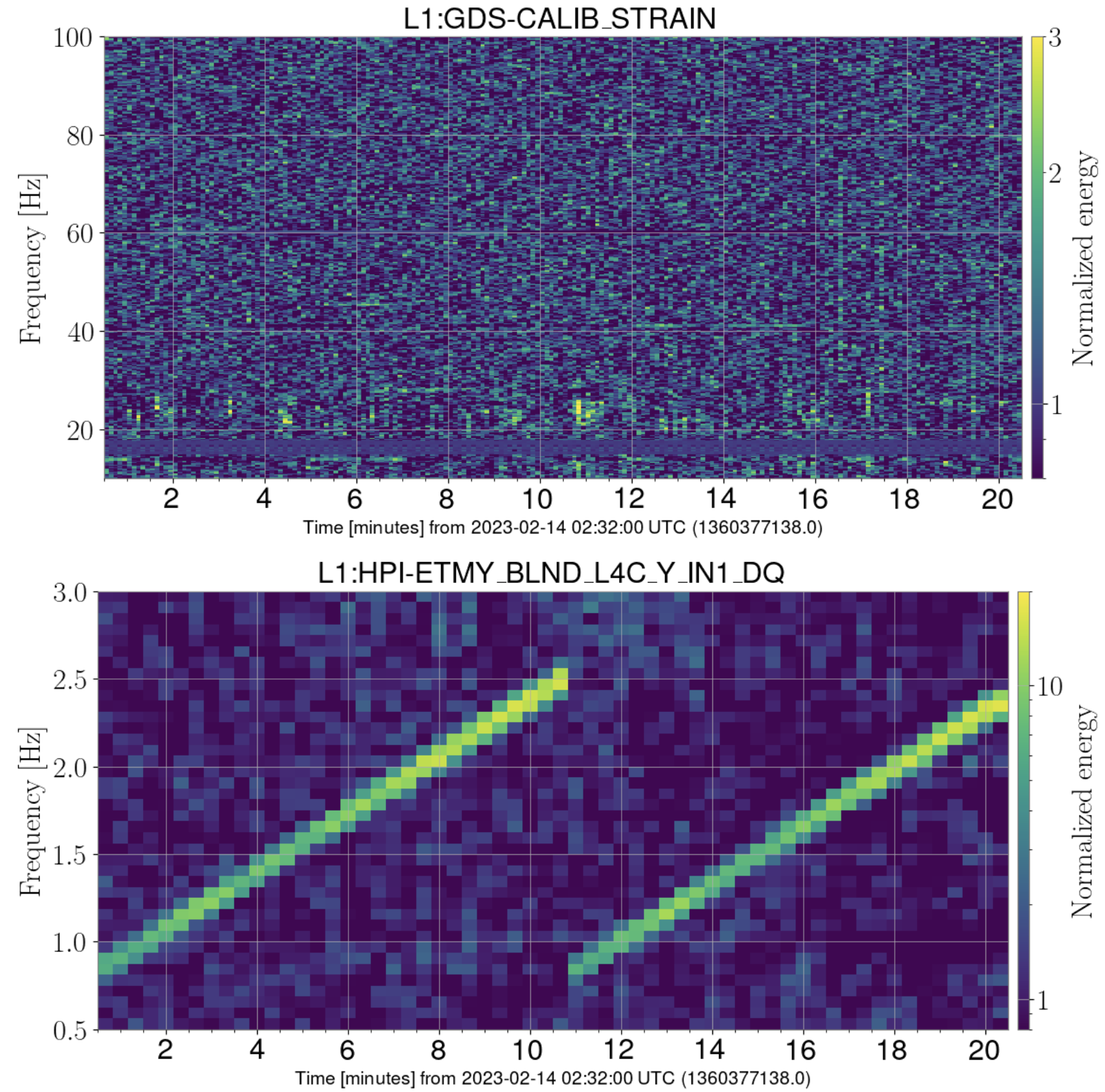}
    \caption{As shown in the left plot in Fig \ref{fig:acb_etmy_inj}, PEM injections identified resonance close to 1.6 Hz at End Y ACB in July 2022. The test was repeated on Feb 14 2023 after fixing the ACB resonances in fall 2022. We once again sweep in the same frequency band but do not observe any response in h(t) this time \cite{alog:etmy_hepi_feb_inj}.}
    \label{fig:acb_etmy_inj_test}
\end{figure}

\section{Noise reduction}\label{section_noise_reduction}
In this section we look at the h(t) data quality at LLO in the next lock period following the commissioning work to fix ACB resonances in fall 2022 \cite{alog:acb_fix}. There are two ways we can check the coupling between motion surrounding the ACB and scatter noise in h(t). The first test is to repeat the sweep injections and compare the h(t) response with the previous injections. The second test is to pick out a period of adverse environmental conditions before and after the fix and compare the h(t) data quality. 
Both of these tests are necessary to check the noise coupling. Since the motion induced during injections may differ from the motion induced by trains or other sources, it is essential to examine the h(t) data quality under both sources of motion: injected and environmental.  

\subsection{Sweep Injections Test}
Earlier, sweep injections in the band $0.8-3~\mathrm{Hz}$ at Corner and Y End station had shown noise coupling to h(t). We repeated these injections in February 2023 and compared the h(t) response with the injections in summer 2022. Figure \ref{fig:acb_etmy_inj_test} shows the comparison for End Y. A similar comparison for ITMY and ITMX injection tests can be found here \cite{alog:ACB_sweep_tests_corner}.

As can be seen from Figure \ref{fig:acb_etmy_inj_test}, repeating the sweep injection after the ACB fix, does not result in any visible noise in h(t) \cite{alog:etmy_hepi_feb_inj}. In the left plot of Fig \ref{fig:acb_etmy_inj}, fringing in the band 20 to 100 Hz can be seen as the sweep goes through $\sim$ 1.6 Hz, no such features are visible in h(t) when the test is repated in Feb 2023. 

\begin{figure}[ht]
    \centering
    \includegraphics[width=0.7\textwidth]{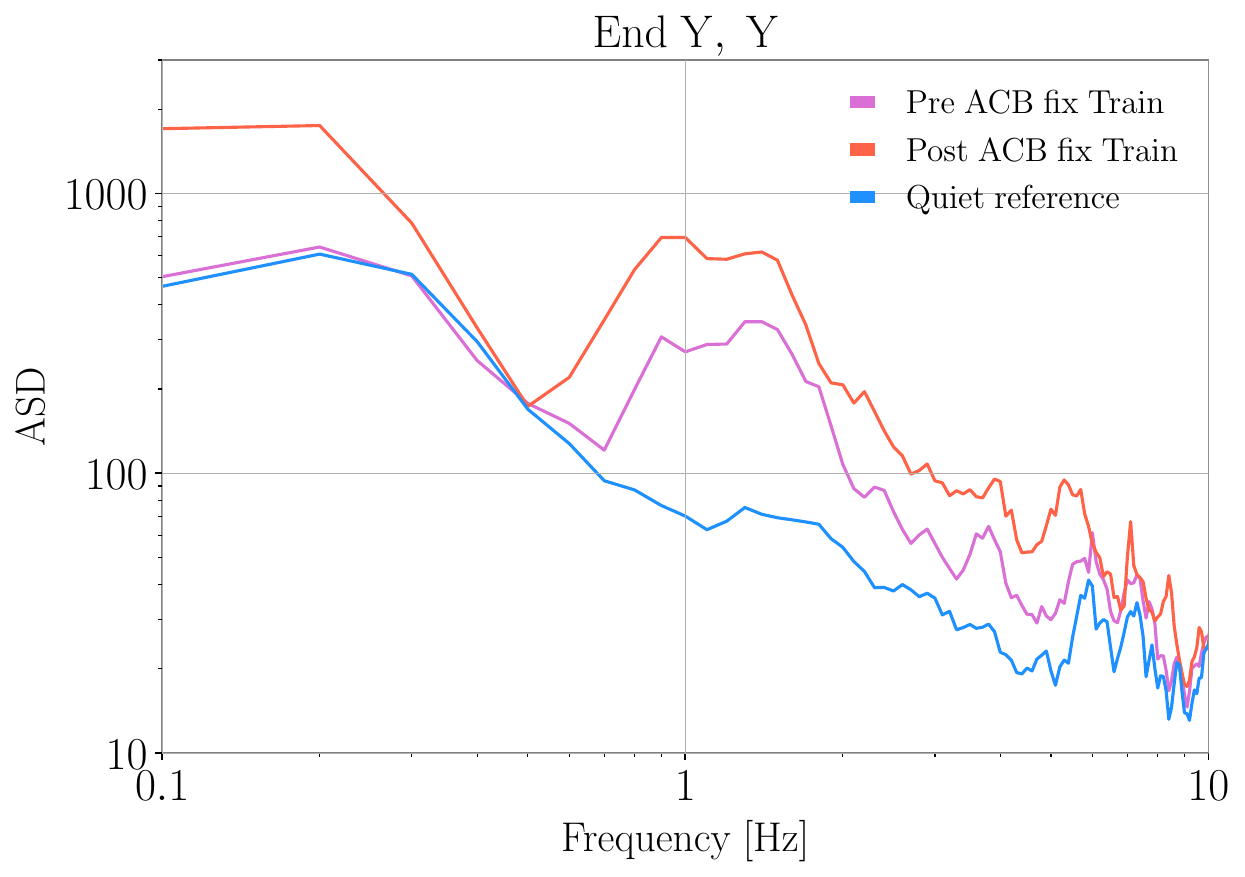}
    \caption{Seismic noise of the two trains before and after the ACB fix. Both trains raise the ASD by a factor of 5 to 10 compared to the quiet time. We look at the h(t) response during these trains in Figure \ref{fig:train_spec_comparison}. }
    \label{fig:asd_may29_mar7}
\end{figure}

\subsection{Trains, logging and other anthropogenic sources}
Ground motion due to trains passing near LLO created noise in h(t) in the O3 and Post O3 data. This coupling between trains and h(t) got worse after O3. The Fast Scatter due to trains in O3 was mostly in the band  $20-60~\mathrm{Hz}$ but in the Post O3 data of November-December 2020 and May 2022, noise due to trains could be seen as high as 200 Hz. Since LIGO is more sensitive in the band $60-200~\mathrm{Hz}$ than the $20-60~\mathrm{Hz}$, post O3 trains were creating $10-40~\mathrm{Mpc}$ range drops for as long as an hour in a day \cite{glanzer2023noise, alog:trains_june_2022, alog:3.3hz_scatter}.

\begin{figure}[ht]
\captionsetup[subfigure]{font=scriptsize,labelfont=scriptsize}
   \centering
    \begin{subfigure}[b]{0.48\textwidth}
        \centering
         \includegraphics[width= \textwidth]{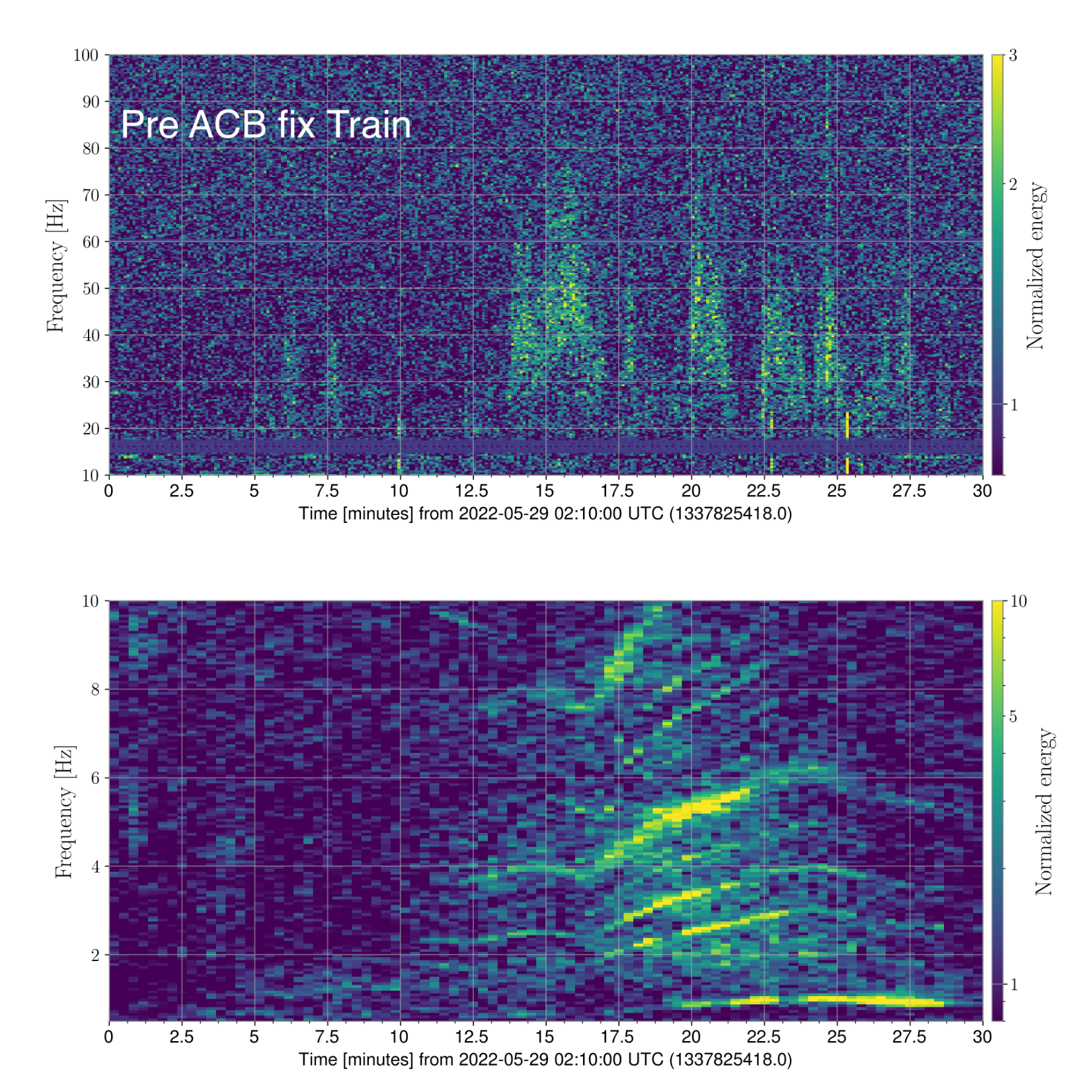}
         \label{fig:train_may29_spec}
    \end{subfigure}
    \hfill
    \begin{subfigure}[b]{0.48\textwidth}
        \centering
         \includegraphics[width =\textwidth]{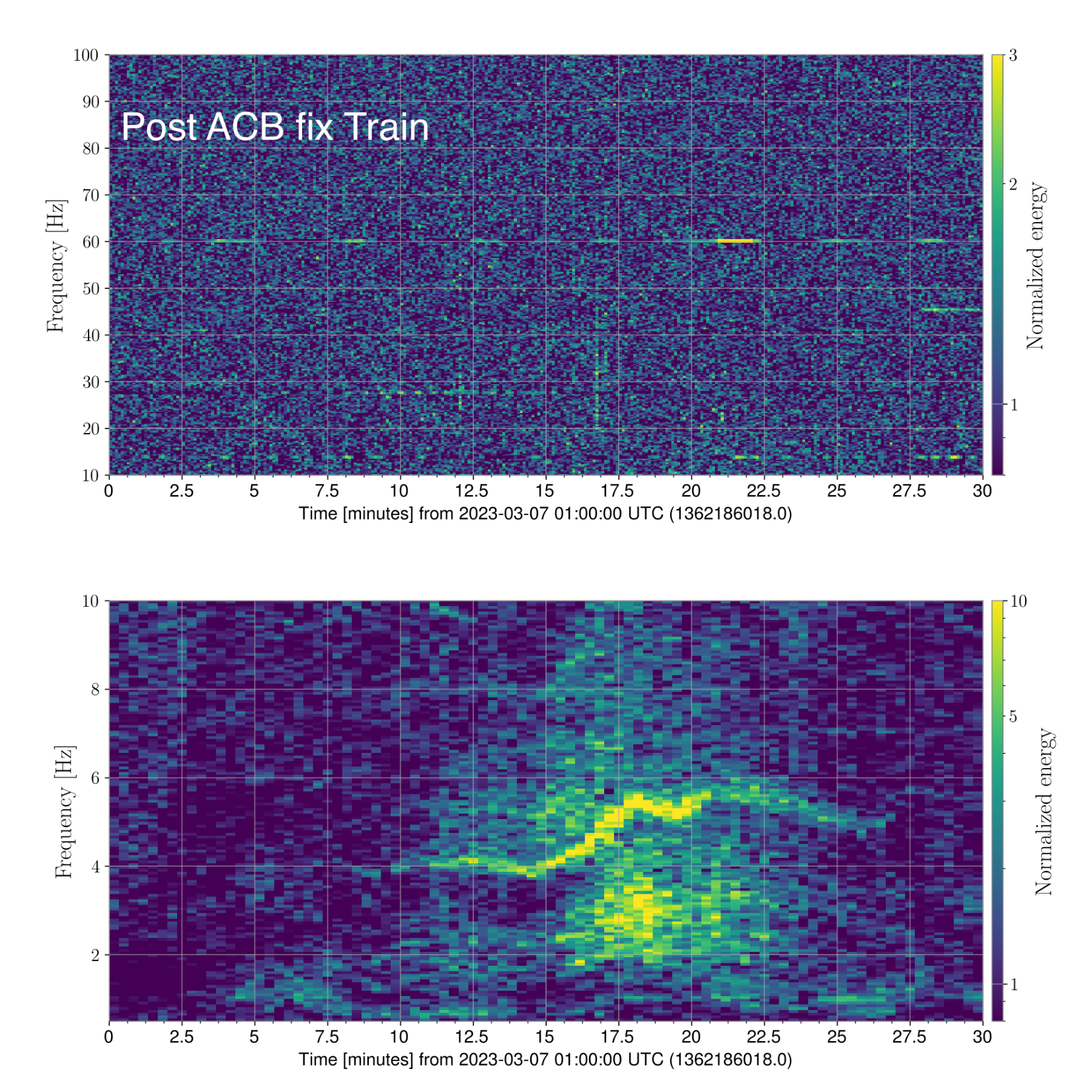}
         \label{fig:train_mar7_spec}    
    \end{subfigure}
    \caption{\emph{Left} \textbf{Pre ACB fix train}: The bottom plot shows the spectrogram of ground motion recorded at End Y along the Y axis. The train appears around the 10 minute mark in the data. The top plot shows h(t) spectrogram for the same duration. Prior to fixing the ACB resonance, trains would create noise in the band $20-200~\mathrm{Hz}$. \emph{Right} \textbf{Post ACB fix train}: We can see the train in the bottom End Y spectrogram but now we do not see any noise in the h(t) spectrogram associated with this train. As can be seen in Figure \ref{fig:asd_may29_mar7}, this train induced more ground motion and yet does not generate significant noise in h(t).}
    \label{fig:train_spec_comparison}
\end{figure}

Figure \ref{fig:asd_may29_mar7} shows the ground motion ASD at the Y End station for two trains. The post ACB fix train (on March 7 2023) is seismically noisier compared to the pre ACB fix train (on May 29 2022) as seen from this figure. We thus expect it to create as much or more noise in h(t) given the same noise coupling. However, this is not what we observe. 
In Figure \ref{fig:train_spec_comparison}, we compare the noise in h(t) at the time of these trains. In the left plot, we can see h(t) noise in the band $20-80~\mathrm{Hz}$ as the pre ACB fix train, shown by the End Y ground motion spectrogram, shakes the ground mostly in $1-10~\mathrm{Hz}$. For the post ACB fix train shown on right, we see no such noise in h(t) \cite{alog:trains_feb_2023}.  We provide more such train noise comparisons in \ref{appen_train}. 

\begin{figure}[ht]
\captionsetup[subfigure]{font=scriptsize,labelfont=scriptsize}
   \centering
    \begin{subfigure}[b]{0.48\textwidth}
        \centering
         \includegraphics[width= \textwidth]{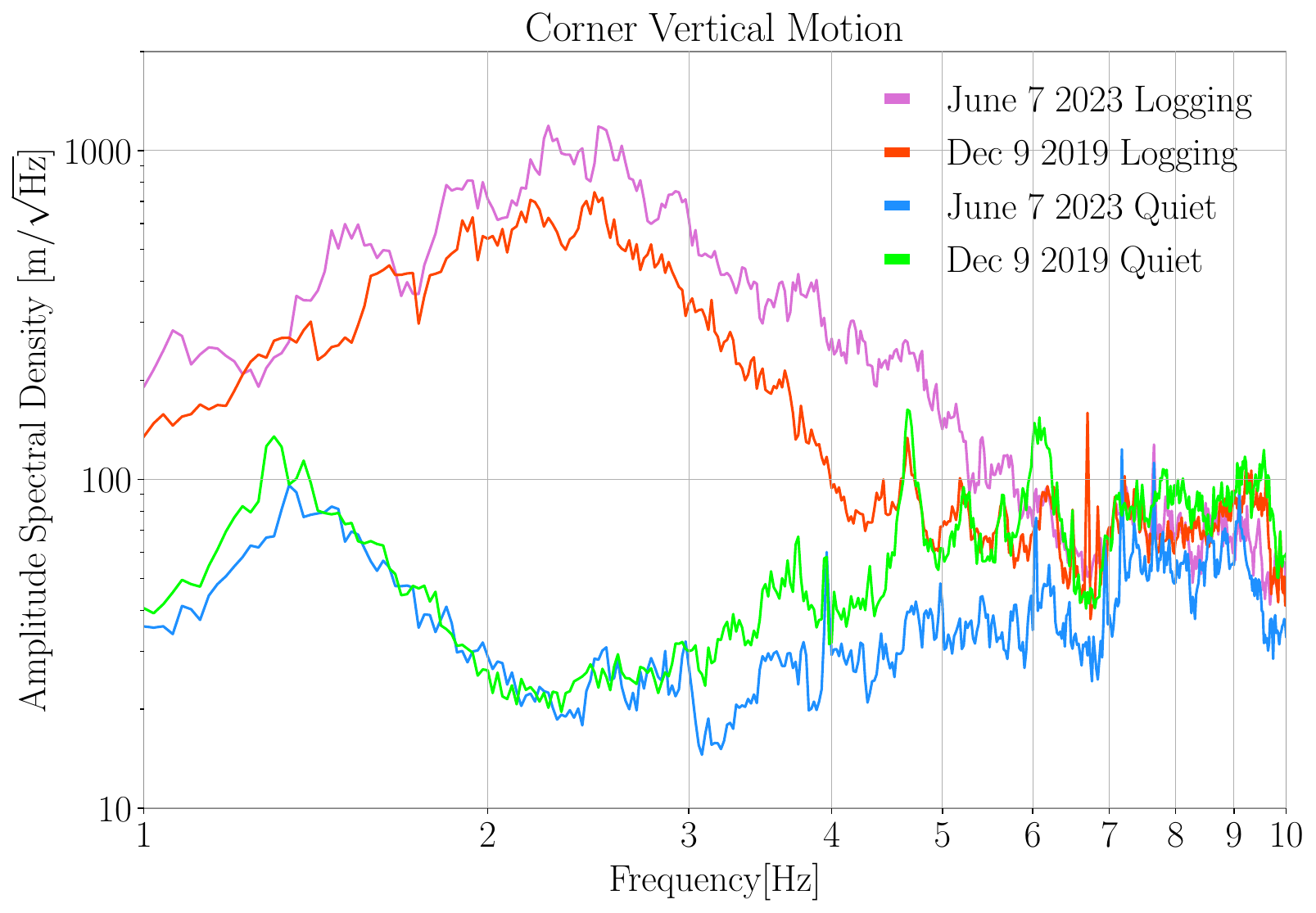}
         \label{fig:asd_june7dec9}
    \end{subfigure}
    \hfill
    \begin{subfigure}[b]{0.48\textwidth}
        \centering
         \includegraphics[width =\textwidth]{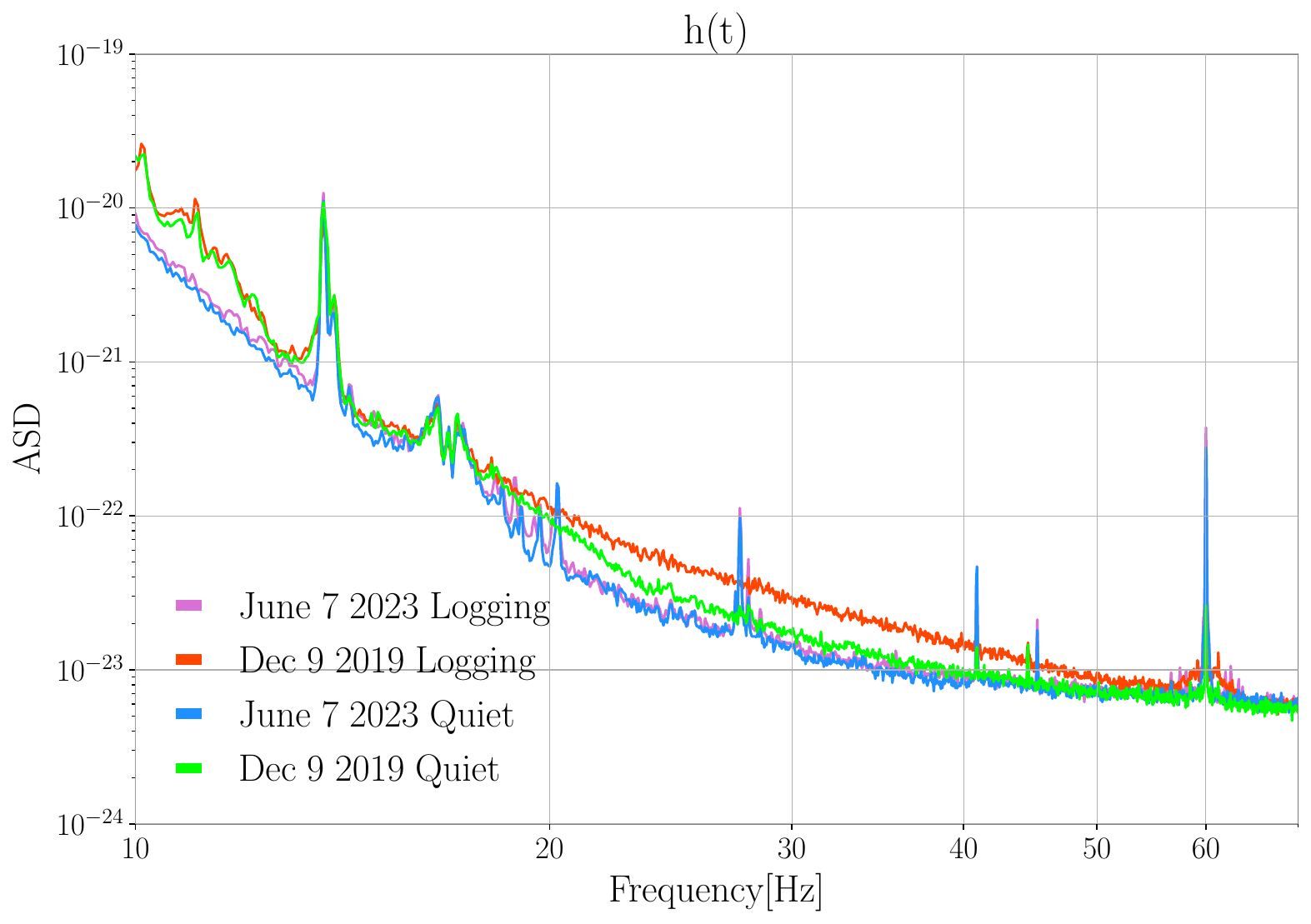}
         \label{fig:darm_asd_june7dec9}    
    \end{subfigure}
    \caption{Comparison of h(t) noise due to logging before and after the ACB resonance fix.
    Here we compare Pre ACB fix Dec 9 2019 with Post ACB fix June 7 2023. Both of these days, logging activities near the Corner station increased the seismic noise mainly in the band $1-6 ~\mathrm{Hz}$. The plots on the \emph{left} show the increase in vertical seismic motion at the Corner station for these days. As shown by the plots on the \emph{right}, the Pre ACB fix logging would create create excess noise in h(t) mainly in the $20-50~\mathrm{Hz}$ band. The Post ACB fix logging however, does not result into any excess noise in h(t).}
    \label{fig:non_train_noise}
\end{figure}

As discussed in Section \ref{section_instrumental_corr}, various activities such as logging, road construction, and trucks on site regularly caused Fast Scatter noise in h(t). Here we compare days with high anthropogenic motion caused by logging sources in O3 and O4. In early Dec 2019, logging work near the Corner station increased the ground motion mostly in the $1-6~\mathrm{Hz}$ band \cite{alog:logging_dec2019}. This increase in seismic noise led to an increase in h(t) noise in the band $20-50~\mathrm{Hz}$. The anthropogenic ground motion and the associated increase in h(t) noise for one such day, Dec 9 2019 is shown in Figure \ref{fig:non_train_noise}. 

In June 2023, a few weeks after the start of the fourth Observing run, logging work began near the Corner station \cite{alog:logging_june2023}. This led to a similar increase in the seismic noise at the Corner station, as shown by the left plot in Figure \ref{fig:non_train_noise}. But this time, we did not observe any excess noise in h(t) data, as shown by the plot on the right. The Corner station seismic noise is visibly higher in the June 2023 logging but it does not lead to any visible noise in h(t) \cite{alog:logging_june2023noise}. This confirms that for non-train anthropogenic ground motion, the noise coupling between seismic noise and h(t) has been reduced significantly. \ref{appen_train} looks at one more such example.


\quad

\begin{figure}[ht]
\captionsetup[subfigure]{font=scriptsize,labelfont=scriptsize}
   \centering
    \begin{subfigure}[b]{0.48\textwidth}
        \centering
         \includegraphics[width= \textwidth]{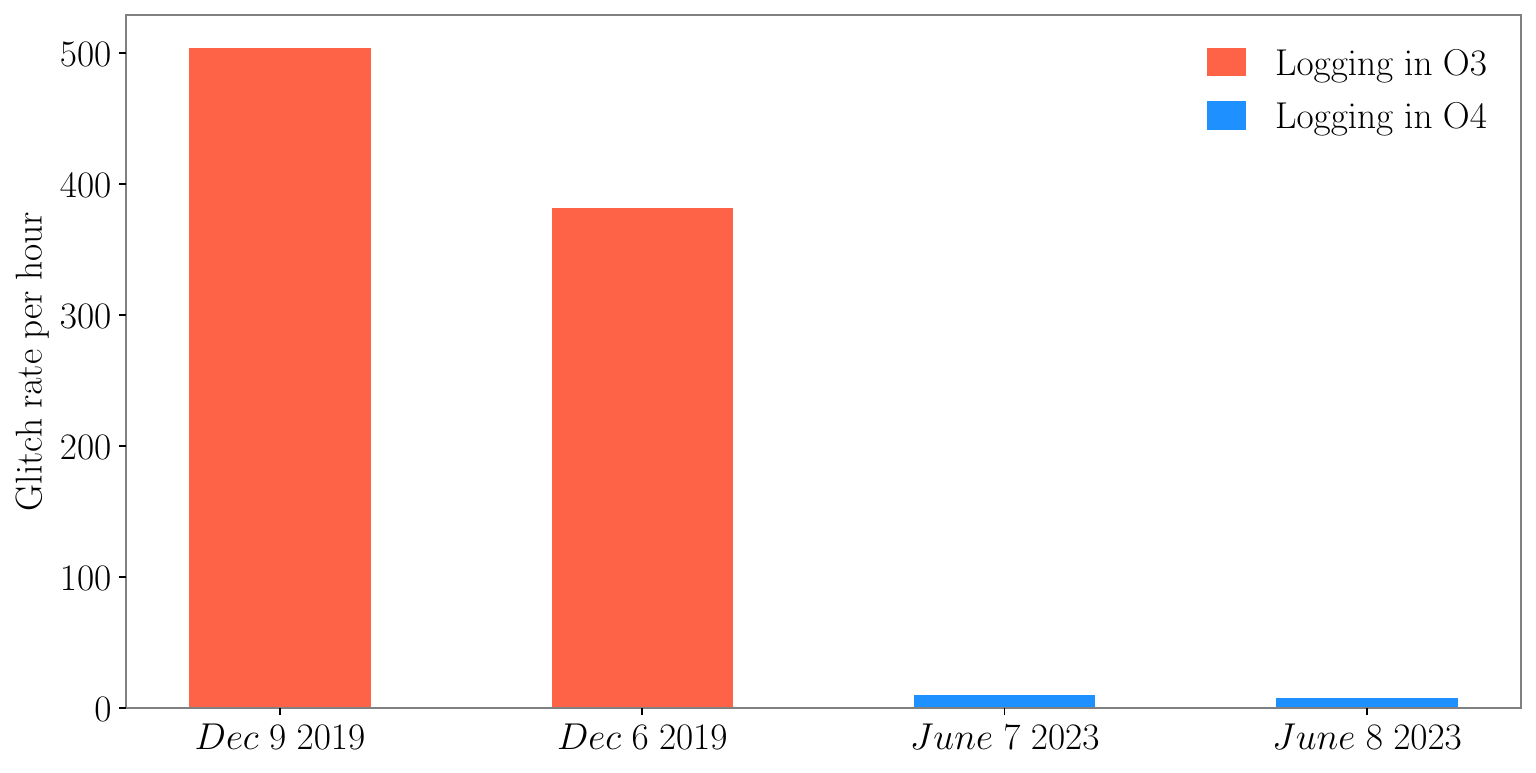}
    \end{subfigure}
    \hfill
    \begin{subfigure}[b]{0.48\textwidth}
        \centering
         \includegraphics[width =\textwidth]{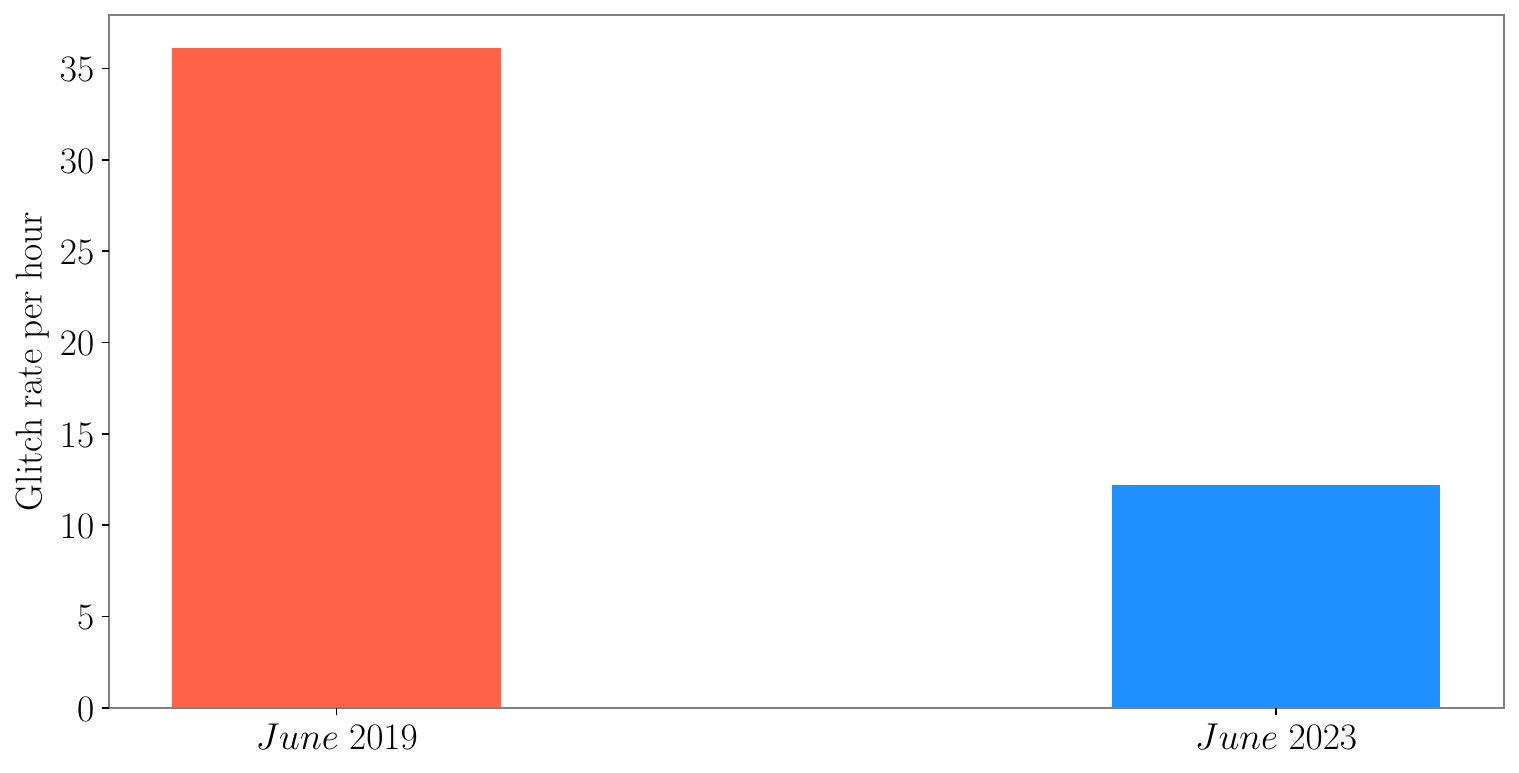}
    \end{subfigure}
    \caption{\emph{Left}: Omicron glitch rate during logging near Corner station at LLO. In O3, an increase in anthropogenic ground motion due to logging would create noise in h(t) and increase the omicron glitch rate substantially. After fixing the ACB resonances, the rate of glitches during logging in O4 is down by a factor of $\sim$ 50. \emph{Right}: Here we compare the glitch rate between June 2019 (O3) and June 2023 (O4). This total glitch rate fell by a factor of $\sim$ 3 in O4. }
    \label{fig:glitch_rate_logging}
\end{figure}


\subsection{Improvement in glitch rate and Binary Neutron Star range}
We compare the rate of omicron glitches during logging in O3 and in O4. Since most Fast Scatter is below 100 Hz, we use this threshold for the comparison. As shown in the left of Figure \ref{fig:glitch_rate_logging}, we have observed a factor of $\sim$ 50 reduction in the glitch rate during logging in O4 as compared to O3. For similar degrees of anthropogenic seismic motion, the rate of transients is greatly reduced. We also compared the total glitch rate between June 2019 (O3) and June 2023 (O4) for a month of omicron triggers in the frequency band  $10-2048~\mathrm{Hz}$. Since the comparison looks at a month long data, it includes different environmental conditions and as such is a better measure of expected glitch rate. Most of this transient noise reduction can be attributed to the ACB fix as Fast Scatter was the most common transient noise during June 2019 and in O3 \cite{glanzer2022data}. This reduction in the rate of transients directly helps in improving the search background used by the search pipelines to identify gravitational waves \cite{Sachdev:2019vvd}. Certain other factors, not currently well understood and not related to the ACB or CB fix are also responsible for a lowered glitch rate, as we have observed transient noise reduction for a wide range of SNR values \cite{alog:glitch_rate_comp_O3_O4}.

The BNS range is the astrophysical distance from which a gravitational wave signal due to the merger of two neutron stars of mass 1.4 $M_\odot$ can be detected with an SNR of 8, averaged over sky locations \cite{davis2021ligo}.
Post-O3 trains were responsible for drops in BNS range as they were creating noise in the higher frequency band compared to trains during O3.
This was especially harmful to the data quality as we may have 1 to 4 trains each day passing near the detector, and each train led to a loss of $10-40~\mathrm{Mpc}$ for $10-20~\mathrm{mins}$. Following the ACB fix, we do not observe the range drops during trains \cite{dcc_LVK_mar_2023_scatter}.

\begin{figure}[ht]
\captionsetup[subfigure]{font=scriptsize,labelfont=scriptsize}
   \centering
    \begin{subfigure}[b]{0.48\textwidth}
        \centering
         \includegraphics[width= \textwidth]{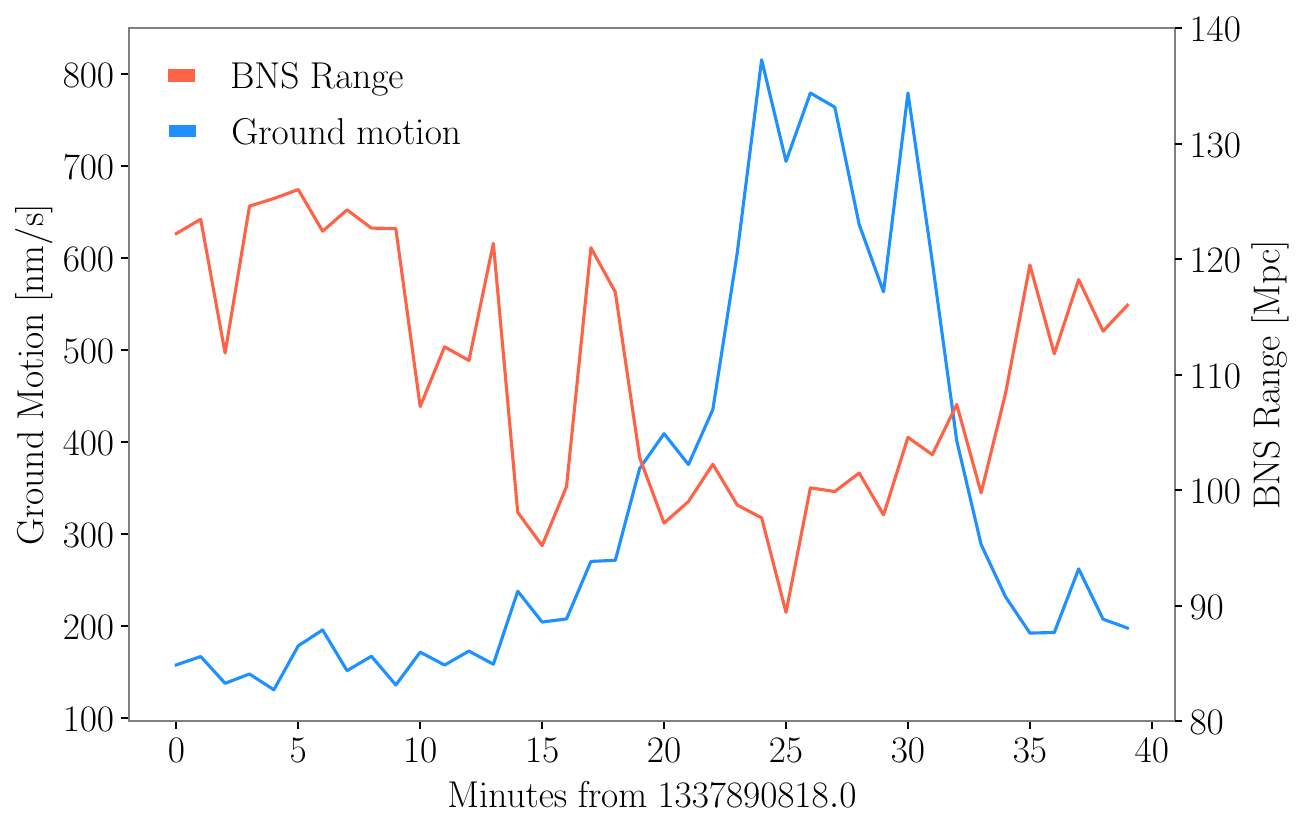}
         \label{fig:range_may29}
    \end{subfigure}
    \hfill
    \begin{subfigure}[b]{0.48\textwidth}
        \centering
         \includegraphics[width =\textwidth]{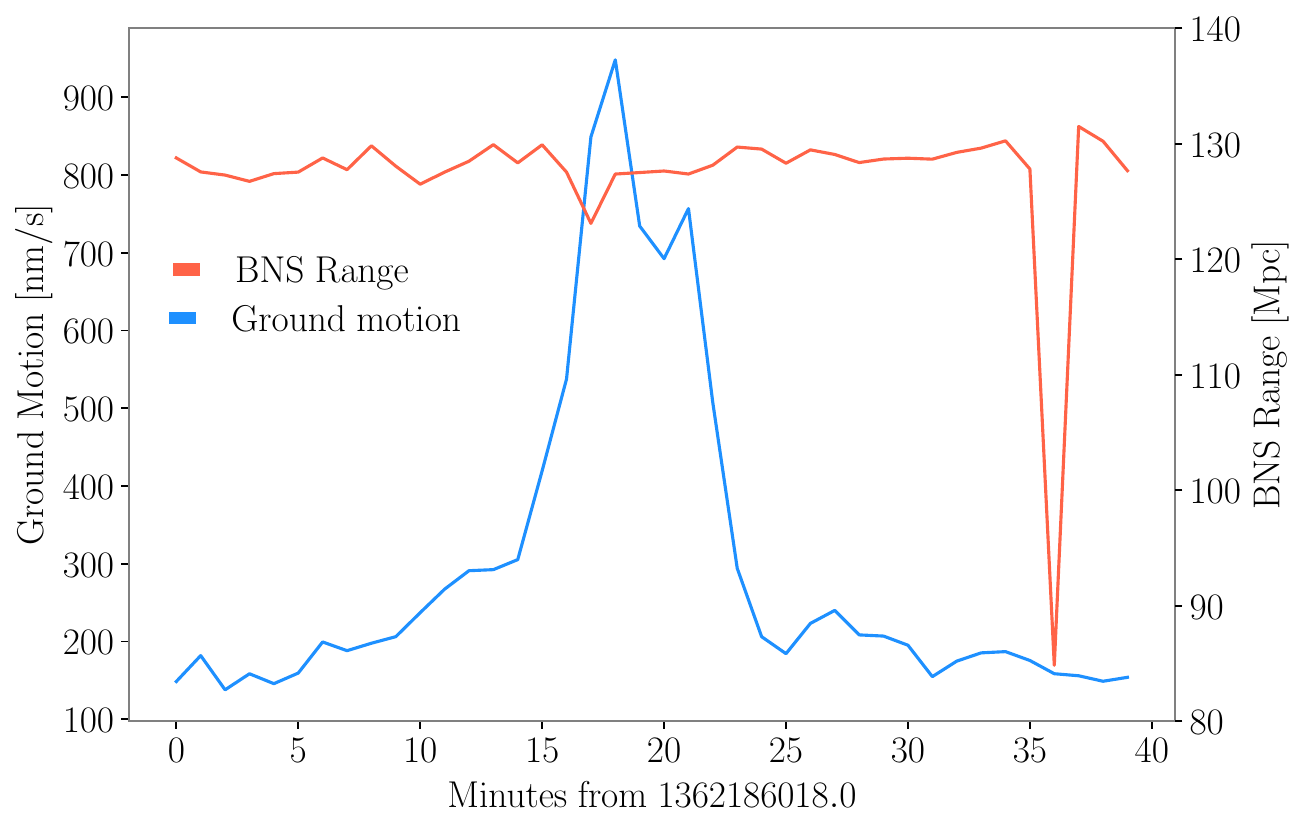}
         \label{fig:range_mar7}    
    \end{subfigure}
    
    \begin{subfigure}[b]{0.48\textwidth}
        \centering
         \includegraphics[width= \textwidth]{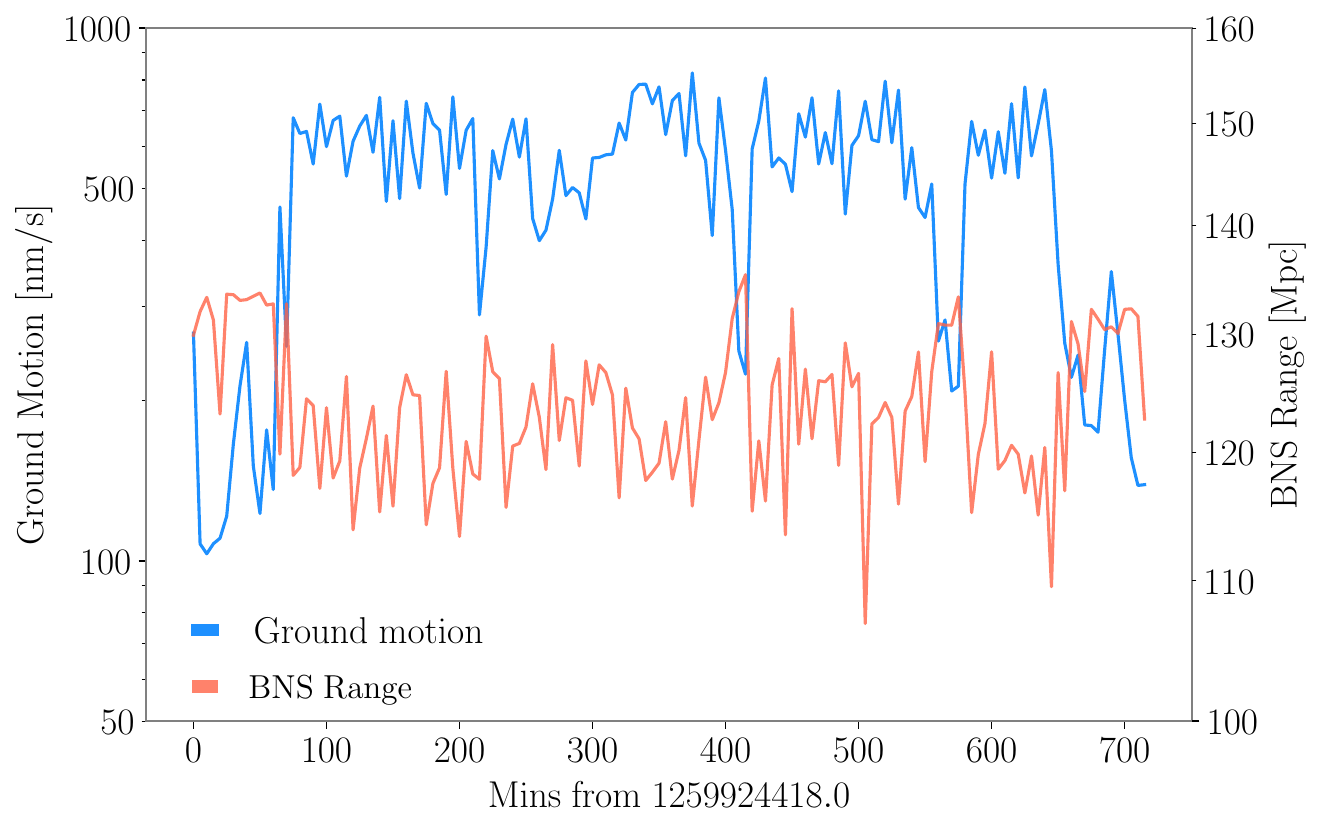}
         \label{fig:range_dec9}
    \end{subfigure}
    \hfill
    \begin{subfigure}[b]{0.48\textwidth}
        \centering
         \includegraphics[width =\textwidth]{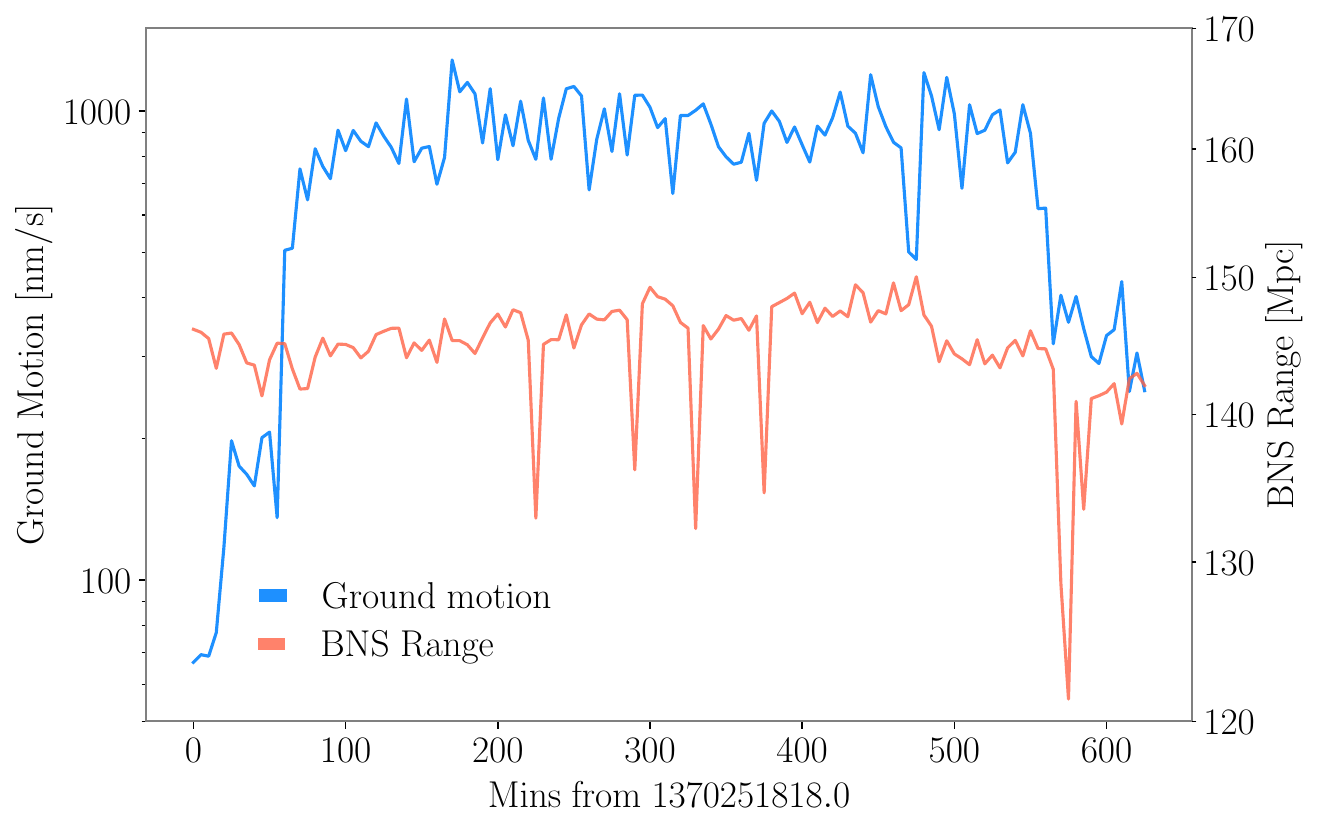}
         \label{fig:range_june8}    
    \end{subfigure}
    \caption{Impact of Trains and logging on BNS range. The plots on the \emph{left} show the reduction in the BNS range during increased ground motion caused by trains (\emph{top}) and logging (\emph{bottom}) before the ACB fix. The plots on the \emph{right} show the improvement in range during similar seismic activities with train on \emph{top} and logging on \emph{bottom} after the ACB was fixed \cite{dcc_LVK_mar_2023_scatter}. The large drops in the top right plot around 35th min mark and in the bottom right plot just before the 600th min mark are due to loud transients and are not related to the ground motion.}
    \label{fig:train_range_comparison}
\end{figure}

The top plots in Figure \ref{fig:train_range_comparison} show the impact of anthropogenic ground motion due to trains on the range before and after the ACB fix. For the train on top left, the range drops coincide with the increase in ground motion in $1-3~\mathrm{Hz}$ band. The plot on the right show a train after the ACB fix and we do not see any significant range drops. Note that the seismic motion induced by these trains is very similar to the trains before the ACB fix. The bottom plots show the range comparision during logging near the detector. This range improvement during trains and logging is a culmination of multiple refinements in the instrument since O3, including but not limited to fixing the ACB and CB resonances.

\newpage
\section{Summary and Conclusion}\label{section_summ_conc}
Noise due to light scattering interferes with our ability to detect gravitational waves, as it introduces additional phase noise into the data. Low frequency seismic motion coupling with high Q resonances in the detector resulted in higher frequency scattering noise observed in h(t) in O3. 
Scattering can be categorized into several types, characterized by the duration of their arches. During O3, several sub-populations of Fast Scatter were observed at Livingston, making it the most common transient noise source. Fast Scatter predominately affected the detector sensitivity between 10 - 100 Hz, but on occasion could reach levels as high as 400 Hz. 

Low frequency ground motion near LLO is quite variable, being influenced by factors such as human activity, earthquakes, and ocean currents in the Gulf of Mexico. This ground motion is a contributing factor to the high levels of scatter present in the O3 h(t) data.  We show through both the modeling of the phase noise and tracking environmental conditions that increases in anthropogenic noise were mainly responsible for 4Hz and 3.3Hz Fast Scatter, whereas increased microseismic noise was mainly responsible for 2Hz Fast Scatter. 

By establishing meaningful correlations between changes in ground motion and models of the phase noise, we were able to identify the surfaces that contribute to light scattering. 
At both sites and across all stations, CB resonances were found using sweep injections. Most of these resonances were damped before O4.
PEM injection tests during 2022 helped reveal the presence of poorly damped 1.6 Hz ACB resonance, which contributed to the 3.3 Hz Fast Scatter observed post O3. Due to the suspension mechanics of these ACBs, the resonant frequency can shift and change the type of noise that appears in h(t). In this way, an ACB resonance at 2 Hz can create both 4 Hz and 2 Hz Fast Scatter depending on the relative amounts of anthropogenic and microseismic ground motion. We showed that the damping of ACB and CB resonances led to major improvements in the transient noise rate,  enhancing the detector sensitivity in pre-O4 data and O4 data.

\ack{
This material is based upon work supported by NSF's LIGO Laboratory which is a major facility fully funded by the National Science Foundation.
LIGO was constructed by the California Institute of Technology and Massachusetts Institute of Technology with funding from the National Science Foundation and operates under Cooperative Agreement PHY-1764464. Advanced LIGO was built under grant No. PHY-0823459. The authors acknowledge support from NSF PHY-2110509. SS acknowledges support from the United States National Science Foundation (NSF) under award PHY-1764464. This work uses the LIGO computing clusters and data from the Advanced LIGO detectors. This document has been assigned LIGO-number LIGO-P2300311.}
\newpage
\appendix
\section{3.3 Hz Fast Scatter}\label{append_3.3Hz}
\begin{figure}[ht]
    \centering
    \includegraphics[width=0.75\textwidth]{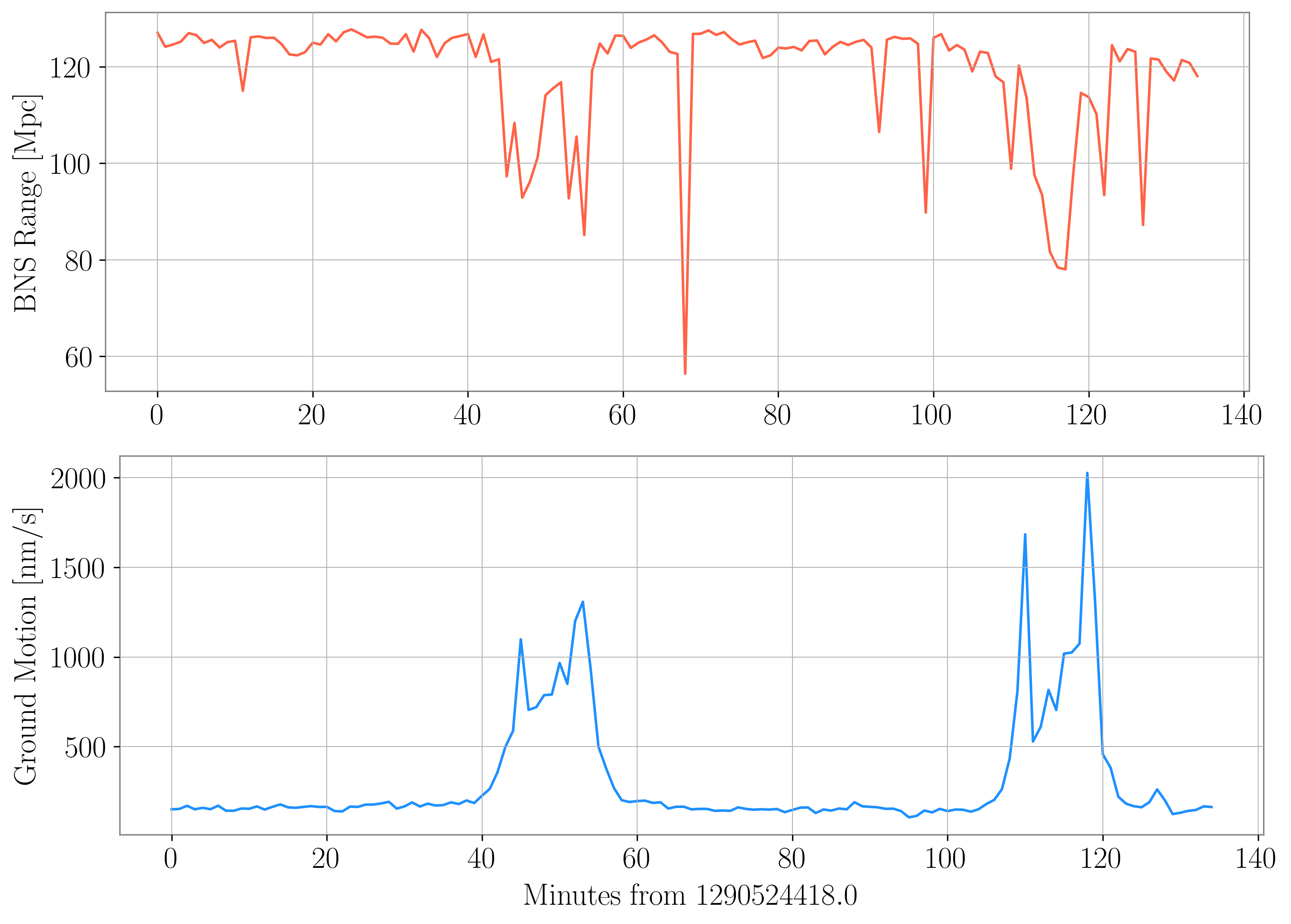}
    \caption{BNS range drops during times of increased ground motion due to trains. The range drops correlate with trains passing near the detector on November 27 2020 from 15:00:00 - 17:20:00. In the absence of ground motion, the BNS range hovers above 120 Mpc. When the two trains pass, the range drops around 40 Mpc.  }
    \label{fig:nov27_gm_range}
\end{figure}

3.3 Hz fast scatter has been observed post O3, with arches separated by $\sim$ 0.3 seconds and an increase in peak frequency as shown in the far right plot in Figure 3. It is similar to 4 Hz scatter in that we have observed trains post O3 that correlate well with an increase in this population of scatter \cite{alog:3.3hz_scatter}. This suggests some change in the noise coupling as during O3, trains would be responsible for 4 Hz Fast Scatter. There were multiple trains on Nov 27 2020, shown in Figure \ref{fig:nov27_gm_range} that gave rise to mainly $3.3$ Hz Fast Scatter in the h(t) data.

\section{Trains and logging noise comparison}\label{appen_train}
Here we look at two more examples of trains and one example of logging before and after the ACB resonance was fixed.

\begin{figure}[ht]
\captionsetup[subfigure]{font=scriptsize,labelfont=scriptsize}
   \centering
    \begin{subfigure}[b]{0.48\textwidth}
        \centering
         \includegraphics[width= \textwidth,height=0.75\textwidth]{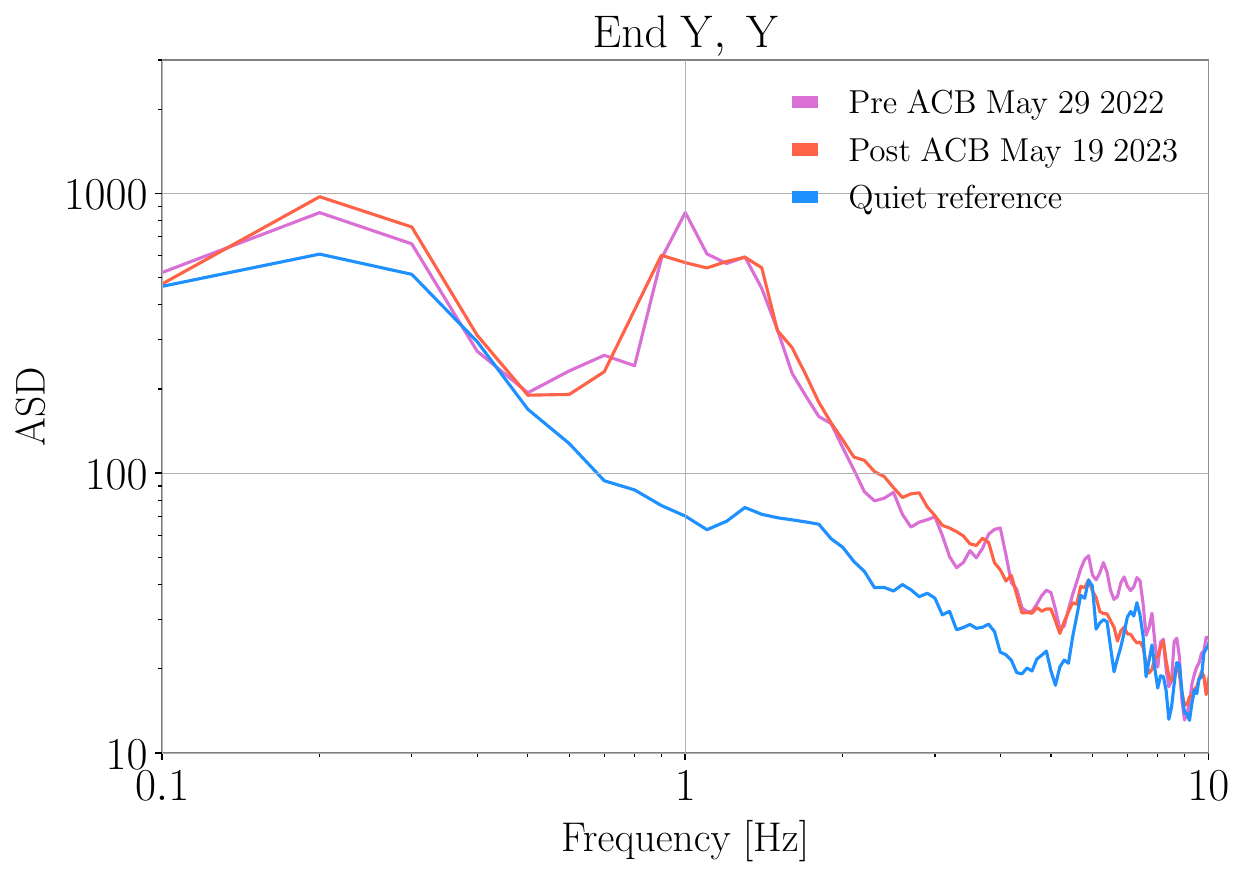}
         \label{fig:train_asd_may29_may19}
    \end{subfigure}
    \hfill
    \begin{subfigure}[b]{0.48\textwidth}
        \centering
         \includegraphics[width =\textwidth]{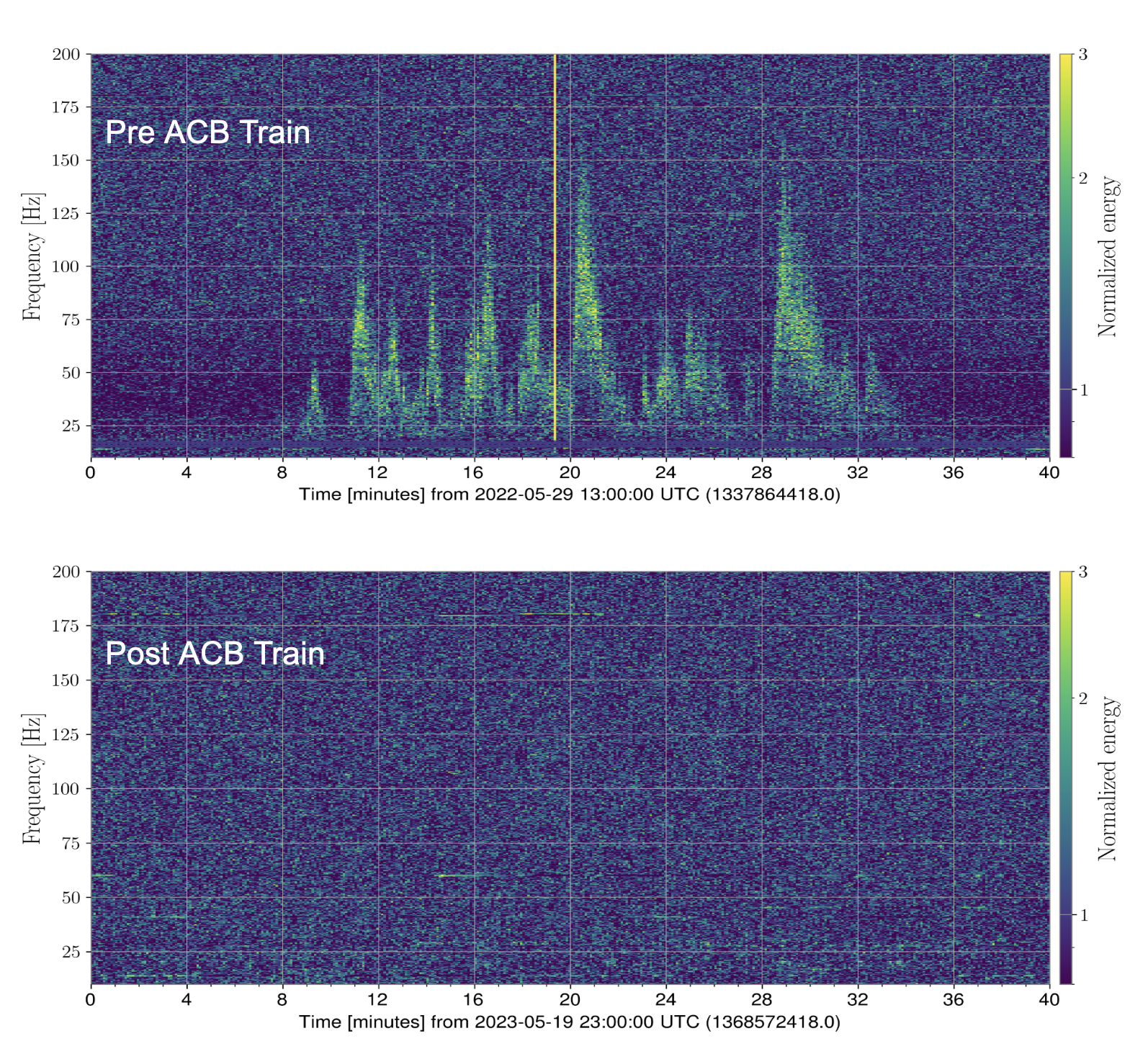}
         \label{fig:train_spec_may29_may19}    
    \end{subfigure}
    \caption{Comparison of trains Pre and Post ACB fix. \emph{Left}: The trains on May 29 2022 and May 19 2023 have a very similar seismic profile, however their impact on h(t) is very different as shown on \emph{Right}. The May 29 2022 train creates scatter noise in h(t) in the band $20-160~\mathrm{Hz}$ while for the train on May 19 2023, the spectrogram looks clean. }
    \label{fig:train_asd_spec_may29_may19}
        
\end{figure}

 In the first example shown in Fig. \ref{fig:train_asd_spec_may29_may19}, we compare a train from May 29 2022 to another train on May 19 2023. As shown in the left plot, the two trains have very similar seismic impact. The first train creates noise in h(t) in the band $20-160~\mathrm{Hz}$ while we do not observe any such noise for the second train as shown by the plot on the right. 

\begin{figure}[ht]
\captionsetup[subfigure]{font=scriptsize,labelfont=scriptsize}
   \centering
    \begin{subfigure}[b]{0.48\textwidth}
        \centering
         \includegraphics[width= \textwidth,height=0.75\textwidth]{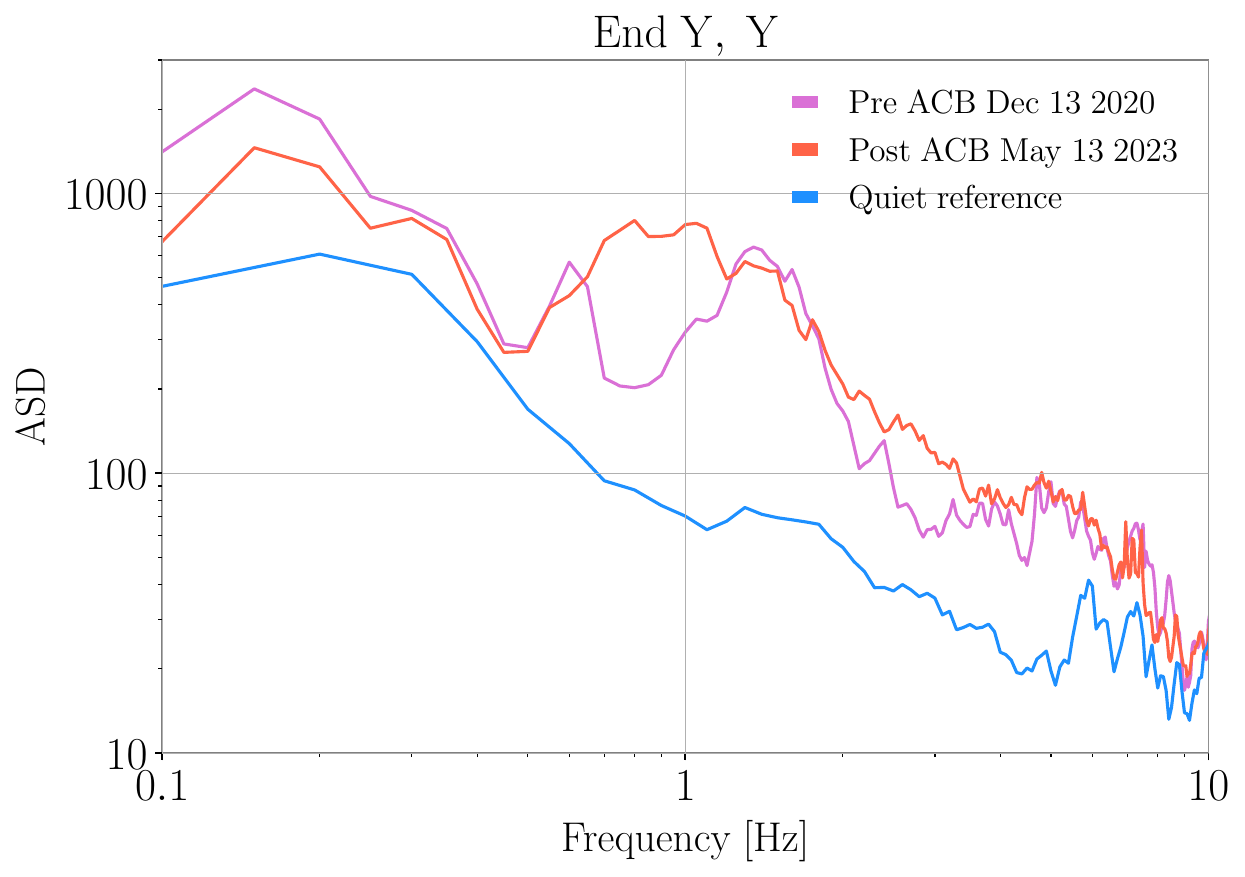}
         \label{fig:train_asd_dec13_may13}
    \end{subfigure}
    \hfill
    \begin{subfigure}[b]{0.48\textwidth}
        \centering
         \includegraphics[width =\textwidth]{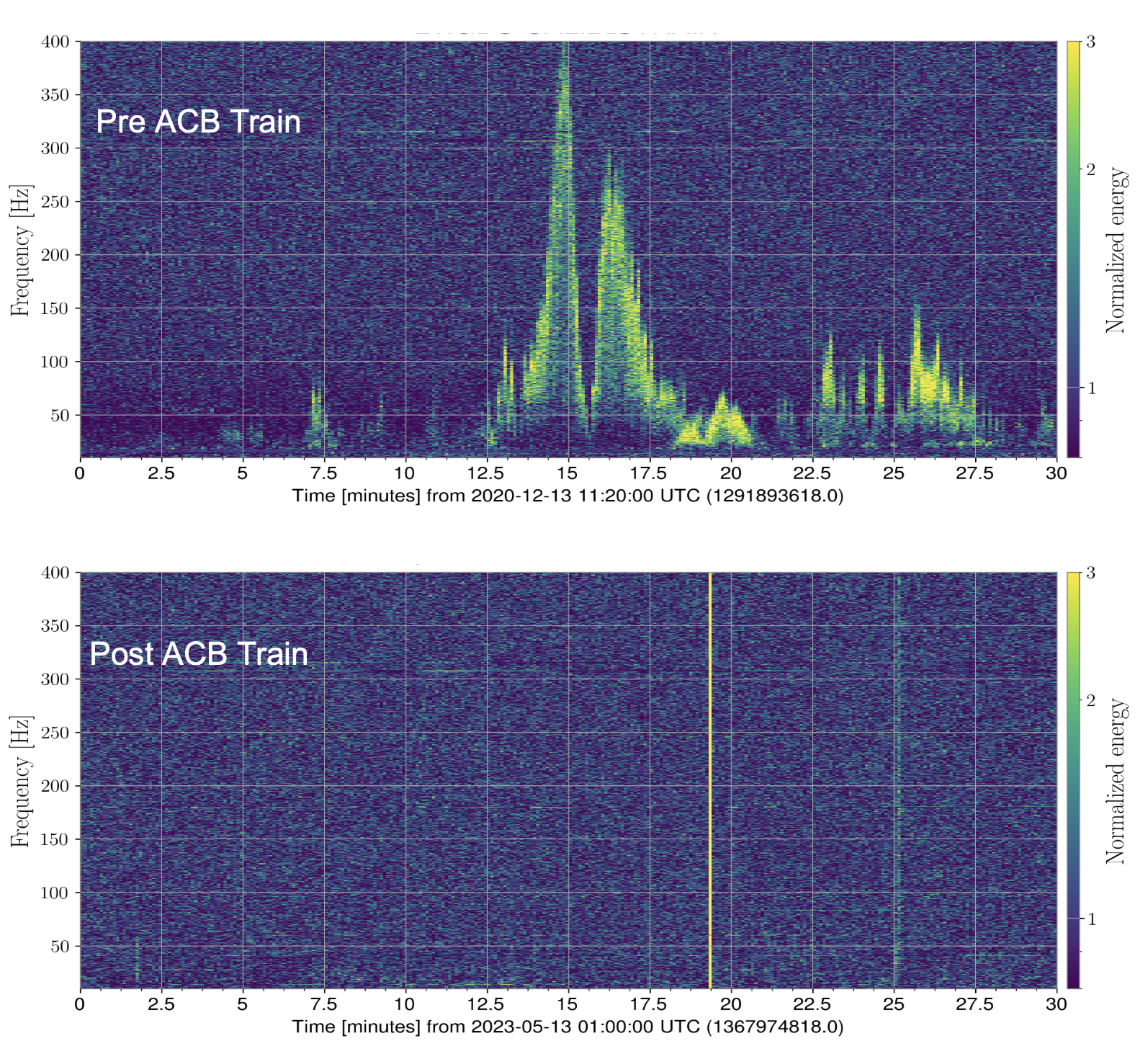}
         \label{fig:train_spec_dec13_may13}    
    \end{subfigure}
    \caption{Comparison of trains Pre and Post ACB fix. \emph{Left}: The trains on Dec 13 2020 and May 13 2023 have a similar seismic profile, with the Post ACB train somewhat noisier seismically. \emph{Right}. The Dec 13 2020 train creates scatter noise in h(t) in the band $20-400~\mathrm{Hz}$ negatively impacting the BNS range. The h(t) spectrogram during the Post ACB fix train on May 13 2023 looks much cleaner as seen here.}
    \label{fig:train_asd_spec_dec13_may13}
\end{figure}

The second comparsion shown in Figure \ref{fig:train_asd_spec_dec13_may13} is between a train on Dec 13 2020 with a train on May 13 2023. Once again, the amplitude spectral density plot on the left shows that the two trains do not differ too much from each other seismically. However their impact on the h(t) is very different. The Pre ACB fix train creates quite a bit of noise reaching as high as $400$ Hz, while the spectrogram for the Post ACB fix train looks very clean.

The next example shows another comparison of noise in DARM during logging activities near the detector Pre and Post ACB resonance fix. 

\begin{figure}[ht]
\captionsetup[subfigure]{font=scriptsize,labelfont=scriptsize}
   \centering
    \begin{subfigure}[b]{0.48\textwidth}
        \centering
         \includegraphics[width= \textwidth]{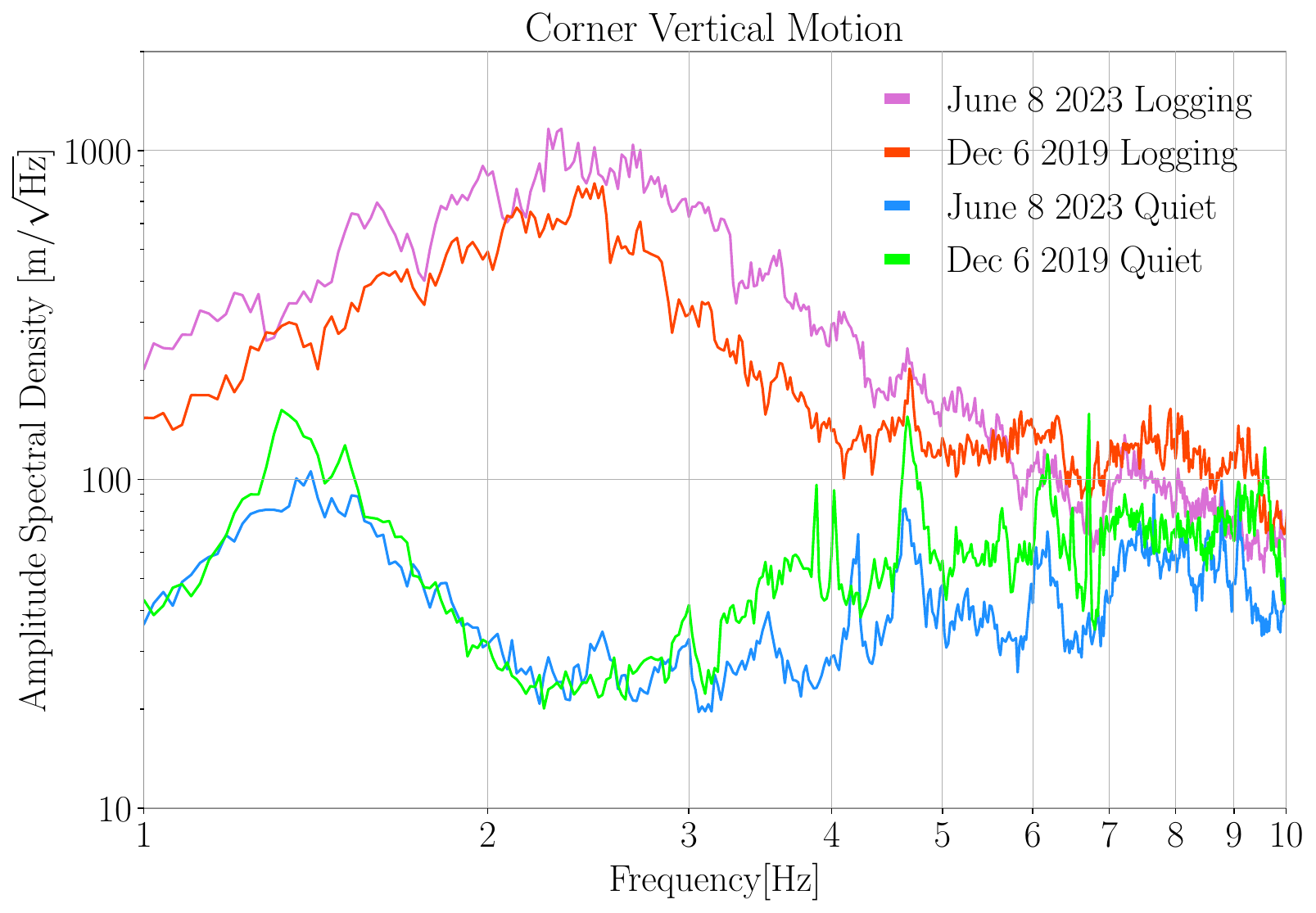}
         \label{fig:asd_may4dec13}
    \end{subfigure}
    \hfill
    \begin{subfigure}[b]{0.48\textwidth}
        \centering
         \includegraphics[width =\textwidth]
         {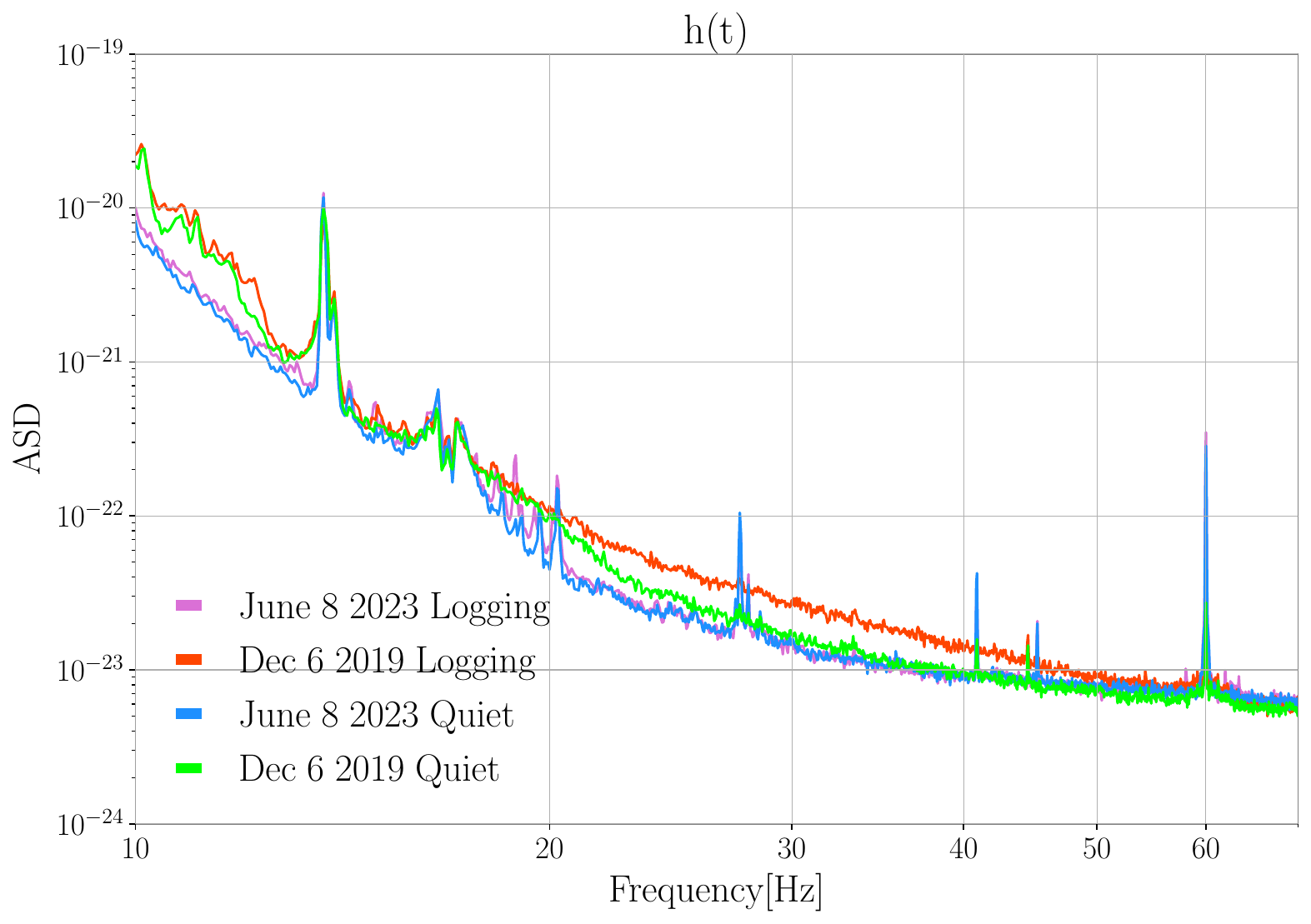}
         \label{fig:darm_asd_may4dec13}    
    \end{subfigure}
    \caption{\emph{Left}: Corner station ground motion on June 8 2023 and Dec 6 2019. Both these days logging activities increased the seismic motion mainly in the band $1-6~\mathrm{Hz}$. Before the ACB fix, this increase in Corner station seismic band motion resulted into an increase in h(t) noise in the band $20-50~\mathrm{Hz}$ as shown on the \emph{right}. Post ACB fix, a similar increase in ground motion no longer creates h(t) noise.}
    \label{fig:non_train_noisemay4}
\end{figure}

\vspace{5em}
\newpage
\bibliographystyle{iopart-num}
\bibliography{sample.bib}

\end{document}

%% file: table_fs_corr.tex
  \begin{tabular}{|c|c|c|c|c|c|} 
\multicolumn{1}{c}{} & \multicolumn{3}{c}{Spearman Coefficient} \\
\cline{1-4}
 \multicolumn{1}{c}{Date}  & \multicolumn{1}{c}{Corner} & \multicolumn{1}{c}{End X}  & \multicolumn{1}{c}{End Y} \\
 \hline
2019-12-11 & $0.55$ & $0.38$ & 0.13 \\ 
\hline
2019-09-22 & $0.29$ & $0.01$ & -0.06 \\
\hline
2019-05-31  & $ 0.58$ & $0.43 $ & 0.30 \\
\hline
\end{tabular}